\documentclass[]{aastex63}
\usepackage{amsmath, bm}
\usepackage{color}
\shorttitle{}
\shortauthors{}
\graphicspath{{./}{figure/}}
\graphicspath{{./figure/}}
\begin{document}

\title{The dynamical structure of the outflows driven by a large-scale magnetic field}

\correspondingauthor{Jia-Wen Li \& Xinwu Cao}
\email{jwlizju@zju.edu.cn, xwcao@zju.edu.cn}

\author[0000-0001-5136-8110]{Jia-Wen Li}
\affiliation{Zhejiang Institute of Modern Physics, Department of Physics, Zhejiang University, \\
38 Zheda Road, Hangzhou 310027, China; jwlizju@zju.edu.cn, xwcao@zju.edu.cn}

\author[0000-0002-2355-3498]{Xinwu Cao}
\affiliation{Zhejiang Institute of Modern Physics, Department of Physics, Zhejiang University, \\
38 Zheda Road, Hangzhou 310027, China; jwlizju@zju.edu.cn, xwcao@zju.edu.cn}
\affiliation{Shanghai Astronomical Observatory, Chinese Academy of Sciences, 80 Nandan Road, Shanghai, 200030, China}
\affiliation{Key Laboratory of Radio Astronomy, Chinese Academy of Sciences, 210008 Nanjing, China}

%% Note that the \and command from previous versions of AASTeX is now
%% depreciated in this version as it is no longer necessary. AASTeX 
%% automatically takes care of all commas and "and"s between authors names.

%% AASTeX 6.31 has the new \collaboration and \nocollaboration commands to
%% provide the collaboration status of a group of authors. These commands 
%% can be used either before or after the list of corresponding authors. The
%% argument for \collaboration is the collaboration identifier. Authors are
%% encouraged to surround collaboration identifiers with ()s. The 
%% \nocollaboration command takes no argument and exists to indicate that
%% the nearby authors are not part of surrounding collaborations.

%% Mark off the abstract in the ``abstract'' environment. 
\begin{abstract}
Large-scale magnetic field is crucial in launching and collimating the jets/outflows. It is found that the magnetic flux can be efficiently transported inward by the fast moving corona above a thin disk. In this work we investigate the dynamical structure of the outflows driven by the large-scale magnetic field advected by the hot corona. With derived large-scale magnetic field, the outflow solution along every field line is obtained by solving a set of magneto-hydrodynamical (MHD) Equations self-consistently with boundary conditions at the upper surface of the corona.  We find that the terminal speeds of the outflows driven from the inner region of the disk are $\sim 0.01c-0.1c$. The temperatures of the outflows at the large distance from the black hole are still as high as several ten keV. The properties of the magnetic outflows derived in this work are roughly consistent with the fast outflows detected in some luminous quasars and X-ray binaries. The total mass loss rate in the outflows from the corona is about $7\%-12\%$ of the mass accretion rate of the disk. The three dimension field geometry, the velocity, temperature and density of the outflows derived in this work can be used for calculating the emergent spectra and their polarization of the accretion disk/corona/outflow systems. Our results may help understand the features of the observed spectra of X-ray binaries and active galactic nuclei. 
\end{abstract}

%% Keywords should appear after the \end{abstract} command. 
%% The AAS Journals now uses Unified Astronomy Thesaurus concepts:
%% https://astrothesaurus.org
%% You will be asked to selected these concepts during the submission process
%% but this old "keyword" functionality is maintained in case authors want
%% to include these concepts in their preprints.
\keywords{Galaxy jets (601); Galaxy accretion disks (562); Accretion (14); Magnetic fields (994)}
% Unified Astronomy Thesaurus concepts: Galaxy jets (601); Galaxy accretion disks (562); Accretion (14); Magnetic fields (994)
% 1.

%% From the front matter, we move on to the body of the paper.
%% Sections are demarcated by \section and \subsection, respectively.
%% Observe the use of the LaTeX \label
%% command after the \subsection to give a symbolic KEY to the
%% subsection for cross-referencing in a \eqref command.
%% You can use LaTeX's \eqref and \label commands to keep track of
%% cross-references to sections, Equations, tables, and figures.
%% That way, if you change the order of any elements, LaTeX will
%% automatically renumber them.
%%
%% We recommend that authors also use the natbib \citep
%% and \citet commands to identify citations.  The citations are
%% tied to the reference list via symbolic KEYs. The KEY corresponds
%% to the KEY in the \bibitem in the reference list below. 
%%%%%%%%%%%%%%%%%%%%%%%%%%%%%%%%%%%%%%%%%%%%%%%%
\section{Introduction} \label{sec:intro}
%%%%%%%%%%%%%%%%%%%%%%%%%%%%%%%%%%%%%%%%%%%%%%%%
Outflows/winds are ubiquitous phenomenon observed in accreting systems with a large dynamical range in black hole masses, ranging from supermassive black hole (SMBH) as the engine of active galactic nuclei (AGNs) to the stellar mass black hole hiding in the galactic binary systems. These outflows/winds often manifest themselves by the blueshifted absorption features in the observed spectra \citep[e.g.,][]{2009ApJ...701..493R,2010A&A...521A..57T,2013MNRAS.430.1102T,2015Natur.519..436T,2015MNRAS.451.4169G,2017MNRAS.469.1553P,2020ApJ...895...37R}, which may be one of the important feedback mechanisms influencing the host galaxies' dynamics, star formation, or even the growth of their central black holes \citep[][]{2005ApJ...620L..79S,2007ARA&A..45..117M,2012ARA&A..50..455F,2017MNRAS.472..949B}. The launching mechanisms of these outflows, however, are still debated. Plausible outflow launching mechanisms include those driven by the radiation force \citep[e.g.,][]{2000ApJ...543..686P,2004ApJ...616..688P,2014ApJ...789...19H,2018ApJ...867..100Y}, the centrifugal magnetic force \citep[e.g.,][]{1982MNRAS.199..883B,1994A&A...287...80C,1994ApJ...429..139C,1997A&A...319..340F,2010ApJ...715..636F,2013ApJ...765..149C,2015ApJ...805...17F,2019ApJ...872..149L,2021ApJ...906..105C,2021ApJ...914...31Y,2021arXiv211010954Y}, and/or the force due to the thermal pressure gradient of the gas \citep[e.g.,][]{1983ApJ...271...70B,2018MNRAS.476.4395B}. The thermally driven winds are often employed to account for a kind of weak and slow outflows launched from very large radius.

The ultra-fast outflows (UFOs) are flowing outward at near-relativistic speeds of $v/c\gtrsim 0.03$, which can be very powerful. Most of them have been observed with the features of highly ionized metallic element in their X-ray spectra, such as $\rm Fe_{XXV}/Fe_{XXVI}$. The ionization parameter of the gas is usually very high $(\rm log\xi \gtrsim 4)$ with H-equivalent column density of $N_{\rm H}\gtrsim 10^{23}\rm cm^2$. The UFOs seem to be ubiquitous across different types of AGNs \citep[e.g.,][]{2009ApJ...701..493R,2010A&A...521A..57T,2010ApJ...719..700T,2011MNRAS.418L..89T,2013MNRAS.430.1102T,2015Natur.519..436T,2015MNRAS.451.4169G,2017MNRAS.469.1553P,2020ApJ...895...37R}. The highly ionized outflows are also observed in some X-ray binaries (XRBs) 
\citep[][]{2006Natur.441..953M,2008ApJ...680.1359M,2012ApJ...759L...6M,2014ApJ...784L...2K,2018MNRAS.479.3978K,2021ApJ...906...11W}. This kind of outflows are quite unlikely driven by the radiation force since the central engines in these systems have strong X-ray emission which ionizes the gas around the disk, and then suppresses the UV line force \citep[][]{1990ApJ...365..321S,2000ApJ...543..686P,2004ApJ...616..688P,2014ApJ...789...19H,2015MNRAS.446..663H}. The magnetic driven outflow mechanism is most probably responsible for the UFOs observed in luminous AGNs or XRBs. \cite{2006Natur.441..953M} reported an X-ray-absorbing disk wind discovered in the stellar-mass black hole binary GRO J16552-40. \cite{2018ApJ...864L..27F} modeled the variations of the Fe K properties of the UFOs in a famous quasar PDS 456. Both of them concluded that the outflows/winds must be driven by the magnetic field. The presence of powerful disk wind/outflows in XRBs and AGNs may suggest strong poloidal magnetic fields threading the accretion disks that accelerate the outflows \citep[e.g.,][]{2006Natur.441..953M,2004MNRAS.355.1105F,2015ApJ...814...87M,2015ApJ...805...17F,2016ApJ...821..104Y,2018ApJ...853...40F, 2018ApJ...852...35K,2020ApJ...904...30M,2021ApJ...906L...2K}. 

The launching and collimating mechanisms of outflows/jets require a large-scale magnetic field. A series of magnetohydro-dynamic simulations investigated the time dependent evolution of the resistive accretion disk and the accompanying Outflows/jets \citep*[e.g.,][]{2002ApJ...581..988C,2004ApJ...601...90C,2007A&A...469..811Z,2012ApJ...757...65S,2013ApJ...767...30B,2014ApJ...796...29S,2015ApJ...814..113S,2016ApJ...825...14S,2018ApJ...861...11S,2018ApJ...857...34Z,2020MNRAS.492.1855M}. In the quasi-steady state, a robust outflow observed in the simulation domain and a large-scale magnetic field is established. The torque exerted by the magnetic field on the disk due to the outflow leading to a tight couple between the outflow and inflow, which haven been verified by the simulations with a prescribed background magnetic field \citep[see the review by][and the references therein]{2007prpl.conf..277P}.

It has been suggested that the magnetic field generated through the dynamo processes in the disk can launch the outflows from the disk \citep[][]{1981MNRAS.195..881P,1981MNRAS.195..897P,1992MNRAS.259..604T,1995ApJ...446..741B,1996MNRAS.281..219T,1998ApJ...500..703R,2014ApJ...796...29S,2020MNRAS.494.3656L}. However, the detailed physics of dynamo generation of a large-scale magnetic field is still quite unclear, though there is evidence of the winds driven by the dynamo generated field in XRBs \citep[][]{2021arXiv210809110C}.    

Alternatively, a large-scale poloidal field can be naturally formed through advection of a weak ordered field threading the ambient gas \citep[][]{1989ASSL..156...99V,1994MNRAS.267..235L,2005ApJ...629..960S}, which has been verified by numerical simulations  \citep[][]{2003ApJ...592.1042I,2013ApJ...767...30B,2014ApJ...784..121S,2018ApJ...857...34Z,2020MNRAS.492.1855M}. The external weak magnetic field threading the outer region of the disk may probably be fed by a companion star or the interstellar medium in AGNs, provided they are somewhat magnetized   \citep[][]{1974Ap&SS..28...45B,1976Ap&SS..42..401B,1989ASSL..156...99V,2005ApJ...629..960S, 2019Natur.573...83Z,2019MNRAS.485.1916C,2020MNRAS.492..223C}. 
The magnetic diffusivity, however, is roughly proportional to the turbulent viscosity, i.e., the Prandtl number $\cal{P}_{\rm m}$ is a constant around unity \citep[][]{1979cmft.book.....P,2009A&A...507...19F,2009ApJ...697.1901G}, which implies a large-scale magnetic field threading a turbulent thin disk must diffuse away rapidly, i.e., the transport of the magnetic flux in a turbulent thin disk is very ineﬀicient \citep*[][]{1994MNRAS.267..235L}. Therefore, the key to generate a strong poloidal magnetic field threading a thin disk is to suppress the outward diffusion of the flux when the field is  accumulated inward.

Different attempts have been made to solve this issue. One of these scenarios suggested that, the structure of a thin disk is significantly altered due to the angular momentum loss through the magnetic outflow, and therefore the radial velocity of the gas in the thin disk is significantly increased. Thus, the external field is sufficiently dragged inward by the fast moving gas in the disk, and a strong large-scale magnetic field can be formed due to the counterbalance of outward flux diffuse by the efficient inward field transport \citep[][]{2013ApJ...765..149C,2014ApJ...786....6L,2014ApJ...788...71L,2019ApJ...872..149L,2020A&A...641A.133S}.  Another possibility is the presence of highly conducting fast inward moving layers above a thin disk that substantially suppress the diffusion of the magnetic flux, which may also lead to a rather strong large-scale magnetic field  \citep[][]{2009ApJ...701..885L,2009ApJ...707..428B,2012MNRAS.424.2097G,2013MNRAS.430..822G}, though the strength of the field is not very strong (limited by the gas pressure of the layers above the disk) \citep[][]{2018MNRAS.473.4268C}. Recently, the time-dependent global numerical simulations suggested that for the disk-corona system the mass accretion occurs mainly in the corona region, of which the magnetic flux is efficiently transported inward, and a quasi-static large-scale magnetic field with field lines strongly inclined towards the disk surface is formed. This field accelerates a fraction of the hot coronal gas into the outflows \citep[e.g.,][]{2018ApJ...857...34Z,2020MNRAS.492.1855M}.

\cite{2010ApJ...715..636F} constructed a semi-analytical model of the magnetized outflows launched from the accretion disk, and it is successfully applied to model the X-ray absorber properties of XRBs and AGNs \citep[][]{2017NatAs...1E..62F,2018ApJ...853...40F}. They employed a self-similar prescription of the outflows in the calculations, while the boundary conditions of the outflows have not been properly considered. However, as pointed in \citet{2016ApJ...818..152B}, the physical properties of the wind base (e.g., the field configuration, the gas temperature and the gas-to-magnetic ratio) are crucial in calculating the wind properties.
Recently, we derived a global magnetic field configuration of an accretion disk-corona system \citep[][]{2021ApJ...909..158L}. We find that the weak external magnetic field can be efficiently transported to the inner region by the fast moving corona above the disk. 
The derived field configurations are consistent with those exhibited in the numerical simulations \citep[e.g.,][]{2018ApJ...857...34Z,2020MNRAS.492.1855M}. In this work, we study the global dynamics of the magnetic outflows/winds driven by the magnetic field based on the field configuration derived in \cite{2021ApJ...909..158L}. The outflow solution is self consistently derived with suitable boundary conditions at the upper surface of the corona. This paper is organized as follows, Section \ref{sec:outflow_srtu} contains the model of field advection and the dynamics of the magnetically driven outflows. The numerical set-up is described in Section \ref{sec:num_set}. The results and discussion of the model calculations are given in Sections \ref{sec:results} and \ref{sec:dissc} respectively.

%%%%%%%%%%%%%%%%%%%%%%%%%%%%%%%%%%%%%%%%%%%%%%%%
\section{model}\label{sec:outflow_srtu}
%%%%%%%%%%%%%%%%%%%%%%%%%%%%%%%%%%%%%%%%%%%%%%%%
A fraction of the gas in the disk or the layer above the disk can be accelerated into the outflows by the rotating large-scale magnetic field threading the disk \citep[][]{1982MNRAS.199..883B}. The dynamics of the outflow is described by a set of MHD Equations, and the outflow solution is available by solving these Equations when the field configuration and the suitable boundary conditions at the base of the outflow are specified \citep[e.g.,][]{1982MNRAS.199..883B,1994A&A...287...80C,2014ApJ...783...51C}.

As introduced in Section \ref{sec:intro}, the large-scale magnetic field can be formed through advection of the external weak field inward by the gas in the disk or the upper layers (e.g., hot corona). In this paper, we explore how the gas is driven by the large-scale magnetic field advected inward by the hot corona. The global configuration of such a magnetic field has been calculated by \citet{2021ApJ...909..158L}, and we adopt the field configuration/strength and the corona properties derived in their paper to calculate the dynamics of the outflows driven from the corona of the disk. In principle,  the disk and outflows are tightly coupled through mass and angular momentum transfers between them by the magnetic field \citep[e.g.,][]{2006A&A...447..813F,2013ApJ...765..149C,2019ApJ...872..149L}. In this work, for simplicity, we 
neglect the impact of the outflows on the dynamics of the disk, which is a reasonable approximation for weak or moderately strong outflows. As pointed by \citet{2018MNRAS.473.4268C}, the strength of the large-scale field dragged inward by the corona is always limited by the gas pressure of the corona, and therefore only outflows with low or moderate power can be formed in this case.   

In this work, We consider steady axisymmetric outflows driven from the hot corona above the disk, and we shall work in cylindrical coordinates ($R, \varphi, z$), i.e., $\partial/\partial t=\partial/\partial \varphi=0$. With the derived global configuration/strength of the field dragged inward by the corona of the disk, we solve a set of MHD Equations, and the dynamics of the outflows is then derived with the properties of the gas (temperature and density) at the surface of the corona as boundary conditions.

%%%%%%%%%%%%%%%%%%%%%%%%%%%%%%%%%%%%%%%%%%%%%%%%
\subsection{Disk structure and magnetic field configuration}
The large-scale field advection in a turbulent thin accretion disk covered by a layer of tenuous gas or a hot corona was studied in \cite{2021ApJ...909..158L}. Here we briefly summarize the model calculations as follows. As discussed, we have not included the mass loss of the outflows in the calculations of the disk structure. The mass accretion rate is assumed to be a constant radially, i.e., $\dot{M}_{\rm acc} = -2\pi R\Sigma v_{R}= \text{const}$, which implies the mass loss rate in the outflows is negligible compared with the accretion rate in the disk. Here $\Sigma \sim 2 \rho H $ is the disk surface density, and $\rho,H$ are the mean disk density and the scale height of the disk, respectively.
For such a steady accretion disk, the momentum equation reads

\begin{equation}\label{eq:momentum}
    \frac{d}{dR}\left( 2\pi R \Sigma v_{R} R^2 \Omega \right) = \frac{d}{dR}\left( 2\pi R \nu \Sigma R^2 \frac{d\Omega}{dR} \right).
\end{equation}
The radial velocity of the gas in the disk can be calculated by integrating Equation (\ref{eq:momentum}), which reads

\begin{equation}\label{eq:vR}
    v_{R}(R,z) = -\frac{3\nu(z)}{2R}=-\frac{\alpha c_{\rm s}^2(z)}{R \Omega_{\rm k}},
\end{equation}
where $c_{\rm s}$ is the isothermal sound speed, $\Omega_{\rm k}$ is the Keplerian angular velocity, and the the $\alpha$ viscosity $\nu=\alpha c_{\rm s} H $ is adopted. The inner boundary condition of the disk is rather uncertain, and the standard thin disk is artificially cut off at the inner stable circular orbit, which leads to singularities of the disk quantities there \citep{1973A&A....24..337S}. To avoid the complexity of this issue, we adopt Equation (\ref{eq:vR}) in our calculations, which is a good approximation except in the region very close the inner edge of the disk.

The temperature of the disks in AGNs is too low to emit hard X-rays observed in AGNs. The power-law hard X-ray spectra of AGN are most likely due to the inverse Compton scattering of soft photons on a population of hot electrons, most probably in the corona above the disk \citep{1979ApJ...229..318G,1991ApJ...380L..51H,1991ApJ...380...84W,1993ApJ...413..507H}. Such a sandwich disk-corona model has been well developed, and it can successfully reproduce main features of the X-ray spectra of accretion disk-corona systems in different scales, i.e., AGNs, X-ray binaries or even ultrluminous X-ray sources \citep[e.g.][]{1998MNRAS.299L..15D, 1999MNRAS.304..809D,2008ApJ...682..212W,2009MNRAS.394..207C,2001MNRAS.328..958M,2002MNRAS.332..165M,2012ApJ...761..109Y,2016ApJ...833...35L,2017ApJ...841...76L,2018MNRAS.477..210Q,2019A&A...628A.135A,2019ApJ...881...34Y,2019MNRAS.487.5335L,2020MNRAS.495.1158C,2020MNRAS.499.1823Z,2020A&A...642A..94M,2021NatCo..12.1025Y,2021ApJ...910..103L,2021RAA....21..199L}. The disk-corona system is usually described as a two-phase accretion flow with its temperature changing sharply from the cold disk to the hot corona in the vertical direction \citep[][]{1991ApJ...380L..51H,1993ApJ...413..507H}, which is confirmed by the elaborate calculation of the vertical disk-coronal structure with detailed vertical energy transportation included \citep*[e.g.,][]{2002ApJ...575..117L}. As done in \cite{2021ApJ...909..158L}, we choose the following function to describe the gas temperature $\Theta(z)$ along $z$ direction,

\begin{equation}\label{eq:Theta_z}
   \Theta(z) = \left( \Theta_{z_{\rm h}} - \Theta_{0} \right) \frac{e^{b\left( 1 - a \right)} +1}{e^{b\left( 1 - a\frac{z}{z_{\rm h}} \right)} +1} + \Theta_{0},
\end{equation}
where the dimensionless gas temperature is defined as

\begin{equation}\label{eq:def_Theta}
    \Theta(z)\equiv \frac{c_{\rm s}^2(z)}{R^2 \Omega_{\rm k}^2},
\end{equation}
$\Theta_{0}$ and $\Theta_{z_{\rm h}}$ are the dimensionless gas temperature at the disk mid-plane and the surface of the corona (i.e., $z=z_{\rm h}$), respectively. We use two parameters $a$ and $b$ to describe the vertical temperature distribution of the disk-corona. Although the vertical temperature distribution of the disk-corona system is described by an artificial function \eqref{eq:Theta_z}, we believe it indeed reflects the basic feature of the two-phase disk-corona system, i.e., the temperature changes sharply in $z$-direction \citep*[cf., Figure 1 in][]{2002ApJ...575..117L}.

The vertical density profile of the disk-corona system at radius $R$ is obtained by integrating the Equation of  hydrostatic equilibrium from $z=0$ to $z=z_{\rm h}$, i.e.,

\begin{equation}\label{eq:int_hydro_equi}
    {\rm ln}\frac{\rho(R,z_{\rm h})}{\rho_0(R,0)} = -{\rm ln} \frac{\Theta_{z_{\rm h}}}{\Theta_{0}} - \int_{0}^{z_{\rm h}} \frac{z}{R^2 \Theta(z)}d z.
\end{equation}
The gas density decreases with the increase of $z$ from disk mid-plane, thus the corona upper surface at radius $R$ (i.e., $z=z_{\rm h}$, hereafter we refer $z=\pm z_{\rm h}$ as the corona surfaces) is defined as the gas density goes down to a certain critical value $\epsilon\rho _0$, i.e.,

\begin{equation}\label{eq:epsison}
    \frac{\rho(R, z_{\rm h})}{\rho _0(R,0)}=\epsilon,
\end{equation}
where $\rho_0(R,0)$ is the gas density at the disk mid-plane. Combining Equations \eqref{eq:Theta_z}, \eqref{eq:int_hydro_equi} and \eqref{eq:epsison}, we obtain the vertical structure of the disk-corona system, i.e., $z_{\rm h}$ and $\rho(R,z)$, when the parameters, $a$, $b$, $\Theta_{0}$, and $\Theta_{z_{\rm h}}$ are specified.

Based on above disk structure, the advection of the magnetic flux by the fast moving corona can be solved with the induction Equation, i.e.,

\begin{equation}\label{eq:induc_B}
\frac{\partial \pmb{B}}{\partial t}=\nabla \times(\pmb{v} \times \pmb{B}-\eta \nabla \times \pmb{B}),
\end{equation}
where $\pmb{v}$ is the velocity of the plasma and $\eta$ is the magnetic diffusivity caused by the plasma turbulent resistivity. We study the outflow magnetically launched from a diffusive viscous disk-corona system. The magnetic diffusivity $\eta$ is related to the magnetic Prandtl number as ${\cal P}_{\rm m} = \eta/\nu_0$, where $\nu_0$ is the effective turbulent viscosity at the disk mid-plane (i.e., $z=0$) and ${\cal P}_{\rm m}$ is the Prandtl number. Under the axisymmetric assumption, the magnetic field can be described by a stream function $R\psi(R,z)$, and the radial component of the induction equation reduces to (see Equation (10) and also the detailed derivation procedures in \citet{1994MNRAS.267..235L})

\begin{equation}\label{eq:induc_psi}
\frac{\partial }{\partial t} \left[ R \psi(R,z) \right] = -v_{R}(R,z)
\frac{\partial }{\partial R}
\left[ R \psi(R,z) \right] - \frac{4 \pi \eta}{c}R J_ \phi(R,z),
\end{equation}
where $v_{R}$ is the radial velocity of the accreting gas and $J_\varphi$ is the azimuthal current density within the disk and corona. In Equation (\ref{eq:induc_psi}), $ R \psi$ is proportional to the magnetic flux threading the disk within radius $R$, the first term on the right-hand side is the flux advected inward by the accretion flow (both disk and corona), while the second term is the outward flux diffusion due to the turbulent resistivity. As summarized in \citet{2009ApJ...697.1901G}, the net rate on which the poloidal magnetic field being transported inward is governed by the relative intensity of the turbulent viscosity and resistivity, if the former is dominant the magnetic field will be dragged inward efficiently, otherwise, the field will diffuse away, i.e., the field advection is rather inefficient.

The stream function $R\psi(R,z)$ is proportional to the magnetic flux within a circle with radius $R$ located at vertical position $z$.  The azimuthal current density $J_ \varphi (R,z)$ is related to the stream function $R\psi(R,z)$ via the Biot-Savart law,

\begin{equation}\label{eq:psi_Rz}
    \psi_{\rm d} (R,z)=\psi (R,z) - \psi _\infty
    =\frac{1}{c} \int_{R_{\rm in}}^{R_{\rm out}} \int_{0}^{2\pi} \int_{-z_{\rm h}}^{z_{\rm h}} \frac{J_\varphi (R',z') \cos \varphi' d\varphi' R' d R' dz'}{\left[ R^2 + R'^{2} + \left( z-z'\right)^{2} - 2R R' \cos \varphi'\right]^{\frac{1}{2}}} ,
\end{equation}
where $c$ is the light speed and $\psi_{\rm d} (R,z)$ is contributed by the current $J_ \varphi (R',z')$ inside the disk and corona. We assume an external uniform vertical weak magnetic field $B_{\rm ext}$ to be advected inward by the disk-corona system, and therefore $\psi _\infty =B_{\rm ext}R/2$.

For a steady disk, i.e., $\partial / \partial t = 0$, the field advection is in balance with the outward diffusion of the magnetic flux, which is described by

\begin{equation}\label{eq:induc_psi_t_0}
 -\frac{\partial }{\partial R}
\left[ R \psi_{\rm d}(R,z) \right] - \frac{4 \pi \eta}{c} \frac{R}{v_{R}(R,z)}  J_ \phi(R,z) = B_{\rm ext}R.
\end{equation}
Differentiating Equation (\ref{eq:psi_Rz}), we have

\begin{equation}\label{eq:psi_Rz_diff}
    \frac{\partial }{\partial R}\left[ R \psi_{\rm d}(R,z) \right]=\frac{1}{c} \int_{R_{\rm in}}^{R_{\rm out}} \int_{0}^{2\pi} \int_{-z_{\rm h}}^{z_{\rm h}} \frac{\left[R'^2 + (z-z')^2 - RR' \cos \varphi' \right]J_\varphi (R',z') \cos \varphi' d\varphi' R' d R' dz'}{\left[ R^2 + R'^{2} + \left( z-z'\right)^{2} - 2R R' \cos \varphi'\right]^{\frac{3}{2}}} ,
\end{equation}
and substituting Equation (\ref{eq:psi_Rz_diff}) into Equation (\ref{eq:induc_psi_t_0}), it can be re-written as a set of linear Equations

\begin{equation}\label{eq:linear}
    -\sum_{ j=1}^{\rm n} \sum_{ l=1}^{\rm m} J_\varphi (R_{ j},z_{l}) P_{i,j,k,l} \Delta R_j \Delta z_l
    -\frac{4\pi \eta}{c} \frac{R_i}{v_R(R_i,z_k)} J_ \varphi (R_i,z_k) = B_{{\rm ext},k} R_i,
\end{equation}
where 

\begin{equation}
P_{i,j,k,l}= {\frac{1}{c}}\int_{0}^{2\pi}{\frac {\left[ R_j^2 + \left( z_k - z_l \right)^2 - R_i R_j {\cos}\varphi'\right]R_j}{\left[ {R_i}^2+{R_j}^2 + \left( z_k - z_l \right)^2 -2R_i{R_j}\cos{\varphi '}\right] ^{\frac{3}{2}}}}\cos \varphi'{d}\varphi ',
\end{equation}
and the subscripts $``_{i,j,k,l}"$ are labeled for the values of each variables at radius $R_i,R_j$ and vertical position $z_k,z_l$. $J_\varphi(R_j,z_l)$ is the current density at position $(R_j, z_l)$, and $B_{{\rm ext},k} R_i = B_{\rm ext} R_i$ ($B_{\rm ext}$ is the strength of the external uniform vertical magnetic field). Solving a set of the linear Equations \eqref{eq:linear} with given disk structure, i.e., the radial velocity profile $v_R(R,z)$ and the disk thickness $z_{\rm h}$, one can obtain the distribution of the current density within the disk, and then the spatial distribution of the magnetic stream function $\psi (R,z)$. The strength and configuration of the large-scale poloidal magnetic field are then calculated by 

\begin{equation}\label{eq:br_bz}
\begin{aligned}
B_{R}(R,z) =& -\frac{\partial}{\partial z} \psi(R,z),\\
B_z(R,z) =& \frac{1}{R} \frac{\partial}{\partial R}\left[ R \psi(R,z) \right].
\end{aligned}
\end{equation}

%%%%%%%%%%%%%%%%%%%%%%%%%%%%%%%%%%%%%%%%%%%%%%%%
\subsection{Dynamics of the outflows}

A set of MHD Equations describing a steady magnetically driven outflows are \citep[][]{1996ASIC..477..249S}

\begin{equation}\label{eq:MHD}
    \begin{aligned}
        \nabla \times \left(\pmb{v}\times\pmb{B}\right)=&0, \quad (\text{induction equation}) \\
      -\nabla P - \rho \nabla \Psi_{\rm g} + \frac{1}{4\pi}\left( \nabla \times \pmb{B} \right)\times \pmb{B}=&  \rho \left( \pmb{v}\cdot\nabla \right)\pmb{v}, \quad\text{(equation of
motion)}\\
    \nabla \cdot \left( \rho \pmb{v} \right)=&0,  \quad\text{(mass conservation)}\\
    \nabla\cdot \pmb{B}=&0,
    \end{aligned}
\end{equation}
where $\Psi_{\rm g}$ is the gravitational potential.
Under the axisymmetric assumption the magnetic field and the outflow velocity can be decomposed into poloidal (i.e., $B_{\rm p},v_{\rm p}$) and toroidal (i.e., $B_{ \varphi},v_{\varphi}$) components. The dynamical structure of such a magnetized outflow is then described by a set of conservation Equations derived from Equation \eqref{eq:MHD}, which reads \citep[see also][for the details ]{1994A&A...287...80C,2014LNP...876.....R} 

\begin{equation}\label{eq:induc_poloi}
    v_{\varphi} - R\Omega = \frac{v_{\rm p}B_{\varphi}}{B_{\rm p}},
\end{equation}

\begin{equation}\label{eq:eqOfMotion_toroi}
    R v_{\varphi} - \frac{RB_{\varphi}B_{\rm p}}{4\pi\rho v_{\rm p}} = R_{\rm A}^2\Omega,
\end{equation}

\begin{equation}\label{eq:mass_cons}
    \kappa = \frac{\rho v_{\rm p}}{B_{\rm p}},
\end{equation}

\begin{equation}\label{eq:Bernoulli}
\frac{1}{2}v_{\rm p}^2 + \frac{1}{2}(v_{\varphi} - R\Omega)^2 + h + \Psi_{\rm eff} \equiv E.
\end{equation}
Here, $\kappa$ and $E$ are constants along each field line (i.e., the streamline of the outflow), $R_{\rm A}$ is the Alfv\'en radius and $\Omega$ is the angular velocity of the field line at the foot-point $R_{z_{\rm h}}$. Equation \eqref{eq:induc_poloi} is derived from the poloidal component of the induction Equation of the outflow, which is usually interpreted as the gas in the outflow always moving along the field line.    
Equations \eqref{eq:eqOfMotion_toroi} and \eqref{eq:mass_cons} are the conservation of the angular momentum and mass along each field line, respectively. Equation \eqref{eq:Bernoulli} is the Bernoulli integral. The specific enthalpy $dh=dP/\rho$ and with the polytropic law, $P=K \rho^\gamma$, we have

\begin{equation}\label{eq:enthalpy}
    h=\frac{\gamma}{\gamma-1}K \rho^{\gamma-1},
\end{equation}
where $\gamma$ is the adiabatic index and $K$ is a constant along each field line. The effective  potential $\Psi_{\rm eff}$ along the field line that co-rotates with the angular velocity $\Omega$ at its foot-point, i.e.,

\begin{equation}\label{eq:eff_poten}
    \Psi_{\rm eff}(R,z) = -\frac{GM_{\rm BH}}{(R^2 + z^2)^{\frac{1}{2}}} - \frac{1}{2} R^2 \Omega^2,
\end{equation}
where $M_{\rm BH}$ is the mass of the central balck hole. Substituting Equations \eqref{eq:induc_poloi}, \eqref{eq:eqOfMotion_toroi}, \eqref{eq:mass_cons}, \eqref{eq:enthalpy} and \eqref{eq:eff_poten} into Equation \eqref{eq:Bernoulli}, we obtain

\begin{equation}
    \frac{B_{\rm p}^2 \rho_{\rm A}}{8\pi \rho^2} + \frac{\Omega^2 \rho_{\rm A}^2 (R^2 - R_{\rm A}^2)^2}{2R^2(\rho - \rho_{\rm A})^2} + \frac{\gamma}{\gamma-1}K \rho^{\gamma-1} + \Psi_{\rm eff} = E,
\end{equation}
which can be re-written in the dimensionless form,

\begin{equation}\label{eq:Bernoulli3}
    \frac{\xi}{\tilde{\rho}^2}\frac{B_{\rm p}^2}{B_{\rm p,z_{\rm h}}^2} + \frac{\omega}{2}\left[ \frac{\tilde{R}^2 - 1}{\tilde{R}(\tilde{\rho} - 1)}\right]^2 + \theta \tilde{\rho}^{\gamma - 1} - \frac{1}{(\tilde{R}^2 + \tilde{z}^2)^{\frac{1}{2}}} - \frac{\omega}{2} \tilde{R}= \tilde{H}(\tilde{R},\tilde{\rho}, \tilde{R}_{z_{\rm h}}, \tilde{\rho}_{z_{\rm h}})=\tilde{E},
\end{equation}
where the dimensionless parameters are defined as 

\begin{equation}
\begin{aligned}\label{eq:dimen_paras}
    \tilde{R} = \frac{R}{R_{\rm A}},  \tilde{\rho} =& \frac{\rho}{\rho_{\rm A}},\tilde{z} = \frac{z}{R_{\rm A}},\tilde{E} = E/\frac{GM_{\rm BH}}{R_{\rm A}},\\ 
    \xi=\frac{\Theta_{z_{\rm h}}}{\beta_{z_{\rm h}}}\frac{\tilde{\rho}_{z_{\rm h}}}{\tilde{R}_{z_{\rm h}}}, \omega=&\frac{\Omega^2}{\Omega_{\rm k}^2}\frac{1}{\tilde{R}^3_{z_{\rm h}}}, \theta=\Theta_{z_{\rm h}}\frac{\gamma}{\gamma - 1}\frac{\tilde{\rho}_{z_{\rm h}}^{1-\gamma}}{\tilde{R}_{z_{\rm h}}}.
\end{aligned}
\end{equation}
The subscripts $``_{\rm A}"$ and $``_{z_{\rm h}}"$ indicates the values of each quantities at the Alfv\'en point and the bottom of the outflow (i.e., the upper surface of the corona, since we assume the magnetized outflow is launched from the upper surface of the corona), respectively.
The angular velocity $\Omega$ of the disk at the field line foot-point can be calculated by 

\begin{equation}
    R_{z_{\rm h}}(\Omega_{\rm k}^2 -  \Omega^2) =\frac{B_{R}(R_{z_{\rm h}})B_{z}(R_{z_{\rm h}})}{2\pi \Sigma},
\end{equation}
where the magnetic force against the gravity is considered in the calculations of rotation velocity of the gas in the corona. The ratio of gas to magnetic pressure at the surface of the corona $\beta_{z_{\rm  h}}=\rho_{z_{\rm h}}c_{{\rm s}, z_{\rm h}}^2/\frac{B_{{\rm p}, z{\rm h}}^2}{8\pi}$. The mass conservation within the disk requires the gas density at disk mid-plane $\rho(R,0) \propto R^{-3/2}$, and then the dependence of $B_{\rm p}$ on radius $R$ is derived by Equation \eqref{eq:induc_psi}, which is indeed determined by the dynamical structure of the disk, thus  $\beta_{z_{\rm  h}}$ is a function of radius $R$. In this work the magnetic flux is mainly transported by the corona, thus in this region gas pressure must dominate over the magnetic pressure, i.e., $\beta_{z_{\rm  h}}\gtrsim 1$.

For a given magnetic field configuration, a physically plausible magnetic outflow should smoothly pass through three critical points along the magnetic field lines, i.e., the slow sonic point, Alfv\'en point and the fast sonic point \citep[][]{1994A&A...287...80C}. At the sonic points the Bernoulli function \eqref{eq:Bernoulli3} satisfies

\begin{equation}\label{eq:Bern_partial}
    \frac{\partial \tilde{H}}{\partial \tilde{R}}\left.\right|_{\rm s,f} = \frac{\partial \tilde{H}}{\partial \tilde{\rho}}\left.\right|_{\rm s,f} = 0.
\end{equation}
Along any field line, $\tilde{E}$ is a constant, we have

\begin{equation}\label{eq:Bern_const}
    \tilde{H}(\tilde{R}_{\rm s},\tilde{\rho}_{\rm s}, \tilde{R}_{z_{\rm h}}, \tilde{\rho}_{z_{\rm h}}) = \tilde{H}(\tilde{R}_{\rm f},\tilde{\rho}_{\rm f}, \tilde{R}_{z_{\rm h}}, \tilde{\rho}_{z_{\rm h}}) = \tilde{H}(\tilde{R}_{z_{\rm h}},\tilde{\rho}_{z_{\rm h}}, \tilde{R}_{z_{\rm h}}, \tilde{\rho}_{z_{\rm h}})=\tilde{E},
\end{equation}
where the subscripts $``_{\rm s}"$ and $``_{\rm f}"$ refer to the slow and fast sonic points respectively. Thus, we have six Equations, i.e., Equations \eqref{eq:Bern_partial} and \eqref{eq:Bern_const}, to describe the dynamics of the outflow driven by the magnetic field if the field configuration/strength is given.  
 
Thus these six Equations can be solved for six variables ($\tilde{R}_{\rm s}, \tilde{\rho}_{\rm s},\tilde{R}_{\rm f}, \tilde{\rho}_{\rm f},\tilde{R}_{z_{\rm h}}, \tilde{\rho}_{z_{\rm h}}$) with the derived large-scale magnetic field strength and configuration, when the temperature and density at the bottom of the outflow are specified as boundary conditions.

%%%%%%%%%%%%%%%%%%%%%%%%%%%%%%%%%%%%%%%%%%%%%%%%
\section{numerical setup}\label{sec:num_set}
%%%%%%%%%%%%%%%%%%%%%%%%%%%%%%%%%%%%%%%%%%%%%%%%
The configuration/strength of the large-scale magnetic field advected in an accretion disk-corona system is available by solving the linear Equation \eqref{eq:linear}, when the values of the model parameters ${\cal P}_{\rm m} $, $\Theta_{0}$, $\Theta_{z_{\rm h}}$ and $\epsilon$ are specified. The disk-corona is set to extend radially between $R_{\rm in }$ and $R_{\rm out}=1000R_{\rm in}$, and  $R_{\rm in} = 6R_{\rm g}$ ($R_{\rm g}=GM/c^2$) is assumed to be the inner boundary of the disk. We adopt $a=1.5$, and $b=30$ in Equation \eqref{eq:Theta_z} to describe a sharp temperature increase from the cold disk to the hot corona in all the calculations. $J_{\varphi}(R,z)$ is discretized on the center of the cell area $\Delta R_j \times \Delta z_l$. In radial direction we set $n$ grid cells distributed logarithmically within the disk area between $R_{\rm in}$ and $R_{\rm out}$, while the $m$ grid cells in $z$-direction within the upper ($z=z_{\rm h}$) and lower $z=-z_{\rm h}$ surface of the corona is adopted. In all of the calculation we adopt $n=100$ and $m=80$, which can achieve a good performance on accuracy. The distribution of the current density $J_{\varphi}(R,z)$ within the disk-corona (i.e., the region $-z_{\rm h}<z<z_{\rm h}$) is available by solving a set of linear Equations \eqref{eq:linear}, then the spacial distribution of the magnetic potential can be calculated similarly to \cite{1994MNRAS.267..235L}. Equation (\ref{eq:psi_Rz}) can be written in a matrix form, which reads

\begin{equation}
    \left(R \psi_d \right)_{i,k} = \sum_{j=1}^{n}\sum_{l=1}^{m}Q_{i,j,k,l} J_\varphi(R_j, z_l) \Delta R_j \Delta z_l,
\end{equation}
$ \left(R \psi_d \right)_{i,k}$ is related to the magnetic potential at position $(R_i, z_k)$ contributed by all currents within the disk and corona. Matrix $Q$ is defined as 

\begin{equation}
    Q_{i,j,k,l} = \frac{1}{c}\frac{4 R_i R_j}{\left[ \left( R_i+R_j\right)^2 +\left( z_k-z_l\right)^2 \right]^{\frac{1}{2}}}\left[ \frac{\left(2-{\rm k}\right)K(\rm k) - 2 \emph{E} ({\rm k})}{\rm k} \right],
\end{equation}
and 

\begin{equation}
    {\rm k} = \frac{4R_i R_j}{\left( R_i+R_j \right)^2 + \left( z_k-z_l \right)^2}.
\end{equation}
Here $K\left({\rm k}\right) = \int_{0}^{\frac{\pi}{2}} \left( 1 - {\rm k} \sin^2 \theta \right)^{-\frac{1}{2}}d \theta$ and $E\left({\rm k}\right) = \int_{0}^{\frac{\pi}{2}} \left( 1 - {\rm k} \sin^2 \theta \right)^{\frac{1}{2}}d \theta$ are the complete elliptic integrals of of the first and second kind, respectively. A singularity problem has been encountered when evaluate the diagonal elements of matrix $Q$, i.e., at grid cells of $R_i=R_j$ and $z_k=z_l$, thus a smooth procedure is applied, i.e.,

\begin{equation}
    Q_{j,j,l,l} = \frac{\lambda}{4}\left[Q_{j,j+\frac{\lambda}{2},l,l}+ Q_{j,j+\lambda,l,l}  + Q_{j,j-\frac{\lambda}{2},l,l}  + Q_{j,j-\lambda,l,l}
    + Q_{j,j,l,l+\frac{\lambda}{2}} + Q_{j,j,l,l+\lambda} +Q_{j,j,l,l -\frac{\lambda}{2}}+ Q_{j,j,l,l-\lambda}
    \right].
\end{equation}
here, $R_{j+\lambda}$ are the inner and outer boundaries of grid cell $(R_j,z_l)$, $R_{j-\lambda}$, and  $z_{l+\lambda}$,  $z_{l-\lambda}$ are the upper and bottom boundaries, respectively. and $\lambda = 1/2$ is adopted in all the calculations. We find that the final results are almost independent of the value of $\lambda$.

The stream function $R\psi$ is proportional to the magnetic flux in the area within radius $R$, which is a constant along the magnetic field line, thus the field configuration is available with derived $\psi(R,z)$. Finally, the distribution of the field strength is calculated with Equations \eqref{eq:br_bz}.

With the derived large-scale field configuration/strength, the dynamical properties of the outflow along a specific field line are available by solving a set of six Equations (i.e. Equations \ref{eq:Bern_partial} and \ref{eq:Bern_const}) numerically when the parameter $\gamma$ and $\beta_{z_{\rm h}}$ are specified.
The obtained outflow solution passes smoothly through the sonic and Alfv\'en points. The dimensionless density of the outflow along the field line can be calculated with the Bernoulli Equation \eqref{eq:Bernoulli3}. Then, we can calculate other physical quantities (e.g., poloidal/azimuthal velocities, and temperature) of the outflow with Equations \eqref{eq:induc_poloi},  \eqref{eq:eqOfMotion_toroi} and \eqref{eq:mass_cons}. Repeat the calculations along every field line, we obtain the global outflow dynamics. 
\begin{deluxetable*}{cccccccccc}
\tablenum{1}
\tablecaption{model parameters in the calculation of the magnetic\\ field configurations and outflow solution\label{tab:para}}
\tablewidth{0pt}
\tablehead{
\colhead{Model} & \colhead{a}& \colhead{b} & \colhead{$\epsilon$} & \colhead{$\alpha$}&  \colhead{$\cal{P}_{\rm m}$}&  \colhead{$\beta_{z_{\rm h}}(R_{\rm in})$}&  \colhead{$\Theta_0$}&  \colhead{$\Theta_{z_{\rm h}}$}&  \colhead{$\gamma$}
}
\startdata
{} & {} &{} & {} & {} &{} &{}&{}&{}&1.4  \\
Case1: & 1.5 &30 & $10^{-5}$ & 0.1 &1 &1&0.005&0.05&{}\\
{} & {} &{} & {} & {} &{} &{}&{}&{}&5/3 \\ 
\hline
Case2: & 1.5 &30 & $10^{-5}$ & 0.1 &1 &1&0.0025&0.025&1.4\\
\enddata
\tablecomments{$a$ and $b$ are adopted to describe a sharp temperature increase from the cold disk to the hot corona in Equation \eqref{eq:Theta_z}. Under the vertical hydro static equilibrium approximation, $\epsilon$ is used in Equation (\ref{eq:epsison}) to define the location of the corona surface. $\beta_{z_{\rm h}}(R_{\rm in})$ is the plasma $\beta$ parameter at the inner boundary of the disk-corona system. $\alpha$ and $\cal{P}_{\rm m}$ are the viscosity parameter and the magnetic Prandtl number, respectively. $\Theta_0$ and $\Theta_{z_{\rm h}}$ are the dimensionless temperature at the disk mid-plane ($z=0$) and the surface of the corona ($z=z_{\rm h}$), which are defined in Equation(\ref{eq:Theta_z}). Basically we adopt $\gamma=1.4$ in the calculations, for comparison we recalculate case1 with $\gamma=5/3$}
\end{deluxetable*}

%%%%%%%%%%%%%%%%%%%%%%%%%%%%%%%%%%%%%%%%%%%%%%%%
\section{Results}\label{sec:results} 
%%%%%%%%%%%%%%%%%%%%%%%%%%%%%%%%%%%%%%%%%%%%%%%%
The vertical temperature structure of the disk-corona system is described by Equation  \eqref{eq:Theta_z}, and the upper surface of the corona is defined at $z=z_{\rm h}$ in Equation \eqref{eq:epsison}. We adopt ${\cal P}_{\rm m} =1$ and $\epsilon=10^{-5}$, $\alpha =0.1$ and $\beta_{z_{\rm h}}(R_{\rm in})=1$ in all the calculations. We carry out the calculations with the temperature of the gas at the disk mid-plane, $\Theta_{0}=0.005$ ($H/R\simeq 0.07$) and $\Theta_{0}=0.0025$ ($H/R\simeq 0.05$), respectively. All the constants and parameters adopted in the calculations are listed in Table \ref{tab:para}.

\subsection{magnetic field configurations}
%%%%Fig B line%%%%
%%%%%%%%%%%%%%%%%%
\begin{figure}
\gridline{\fig{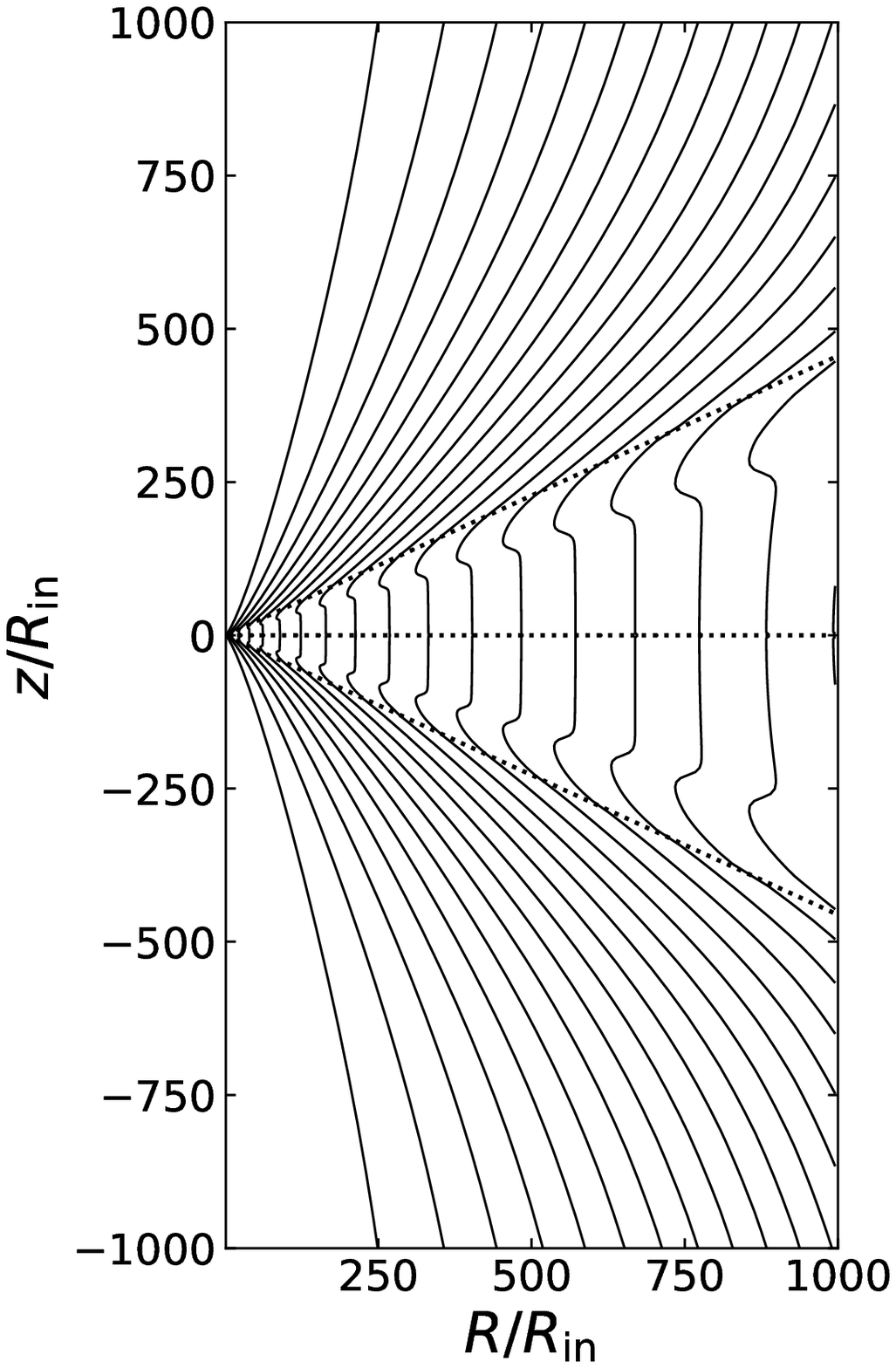}{0.4\textwidth}{}
 \fig{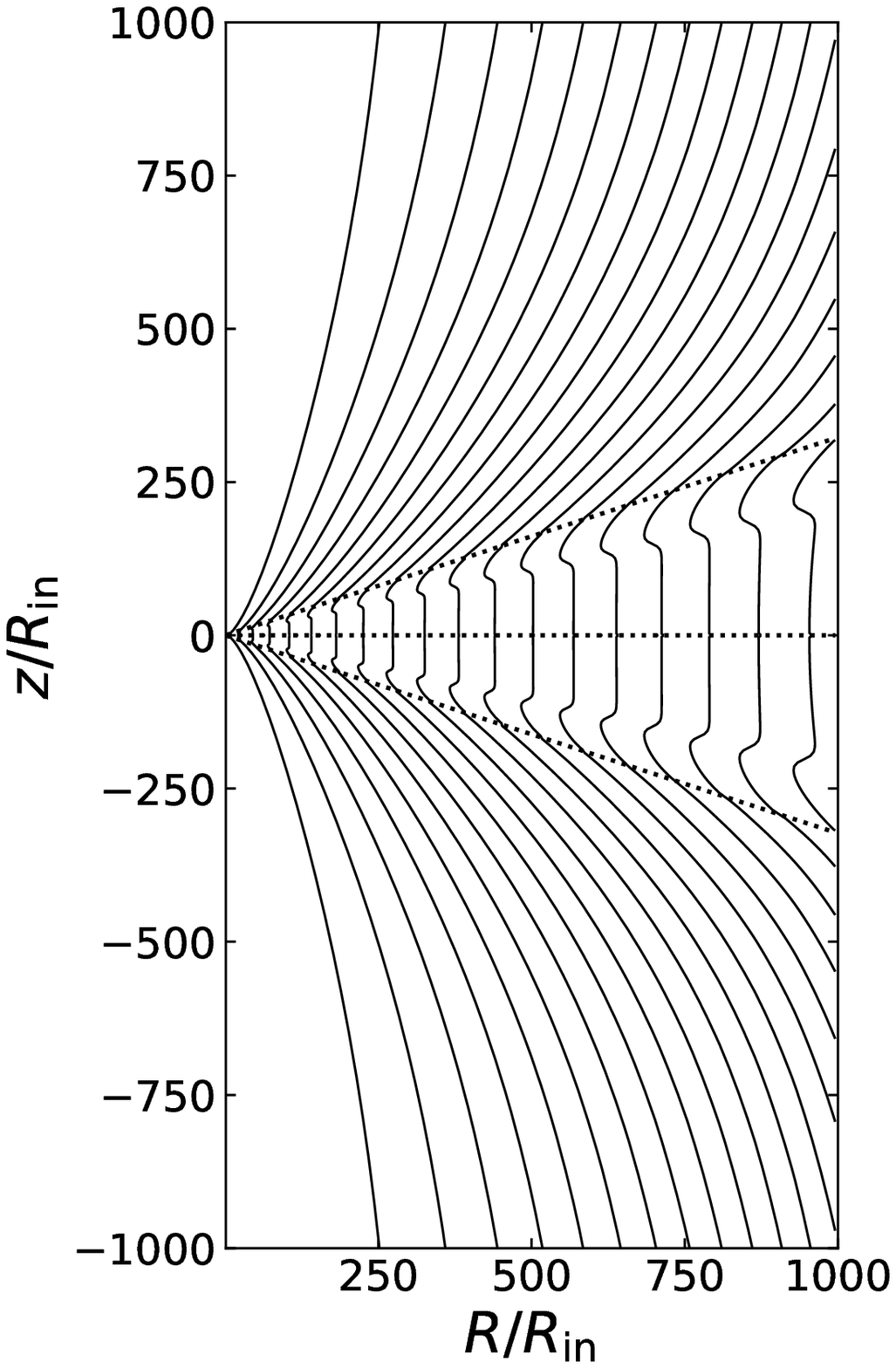}{0.4\textwidth}{}}
\caption{Large-scale poloidal magnetic field configurations. The solid curves are the poloidal magnetic field lines, while the dotted lines show the location of the upper corona surface $z=\pm z_{\rm h}$ (which are derived by Equation \eqref{eq:epsison}) and the disk mid-plane, respectively. Left panel is calculated for case 1: $\Theta_{0}=0.005, \Theta_{z_{\rm h}}=0.05$ and the derived aspect ratio $z_{\rm h}/R = 0.456$. Right panel is calculated for case 2: $\Theta_{0}=0.0025, \Theta_{z_{\rm h}}=0.025$, and the derived aspect ratio $z_{\rm h}/R = 0.323$. $R_{\rm in} = 6R_{\rm g}$ is the inner radius of the disk, and the adiabatic index $\gamma=1.4$ is adopted.
\label{fig:B_line}}
\end{figure}

%%%%%%Fig Bz%%%%%%
%%%%%%%%%%%%%%%%%%
\begin{figure}
    \centering
    \includegraphics[width=0.45\textwidth]{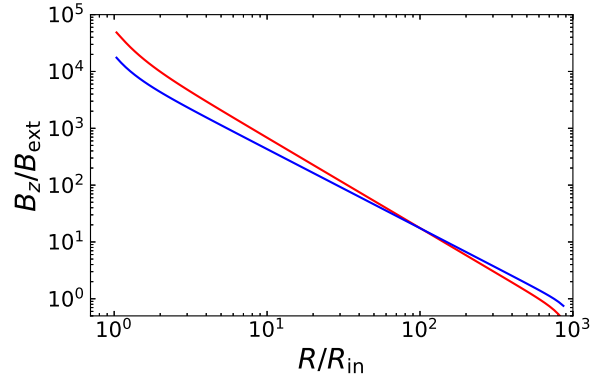}
    \caption{The strength of the vertical component of the large-scale magnetic field, while $B_{\rm ext}$ is the external magnetic field strength. The red line is calculated for case 1: $\Theta_{0}=0.005, \Theta_{z_{\rm h}}=0.05$, while blue line is for case 2: $\Theta_{0}=0.0025, \Theta_{z_{\rm h}}=0.025$
    }
    \label{fig:Bz}
\end{figure}

The configurations of the large-scale poloidal magnetic field advected by the disk-corona are plotted in Figure \ref{fig:B_line}, in which two sets of the parameters are adopted, i.e., case 1: $\Theta_{0}=0.005, \Theta_{z_{\rm h}}=0.05$, and case 2: $\Theta_{0}=0.0025, \Theta_{z_{\rm h}}=0.025$), with an adiabatic index $\gamma=1.4$. The inclination angle of the field line with respect to the upper surface of the corona (i.e., $z=z_{\rm h}$) decreases with increasing field line foot point radius. In this case, the azimuthal field component would be substantially higher than poloidal component not far from the upper surface of the corona, which implies that our assumption of a potential field above the corona is no longer valid. Thus, our calculations will be terminated at a certain radius, and we refer to the region within this radius as the outflow region hereafter. We will  discuss this issue further in Section \ref{sec:dissc}. 

In the region close to the disk mid-plane, the radial velocity is relatively small (see the light blue region of Figure \ref{fig:vp}), the resultant field lines are almost perpendicular to the disk mid-plane (see Figure \ref{fig:B_line}). It indicates that the magnetic diffusion is substantially suppressed in the cold disk even for a strong magnetic field due to a small value of $\nabla \times \pmb{B}$.

The radial profiles of the vertical field component are depicted in Figure \ref{fig:Bz}. The vertical magnetic field strength in the inner region of the disk is increased by several orders of magnitude of the external field strength, which indicates that the magnetic flux is efficiently transported to the inner region of the disk by the hot corona. Due to the presence of the fast moving corona above the disk, the magnetic flux is efficiently transported to the inner region of the disk, which results in a steep slope of the magnetic field strength as a function of radius (see Figure \ref{fig:Bz}), while it is almost flat for a traditional viscous thin disk, i.e., the external field is almost not amplified in a conventional thin disk \citep[see the black line in Figure 6 of][for comparison]{2021ApJ...909..158L}.

\subsection{outflow dynamics}
\label{sec:dynamics}
\begin{figure}
    \gridline{
    \fig{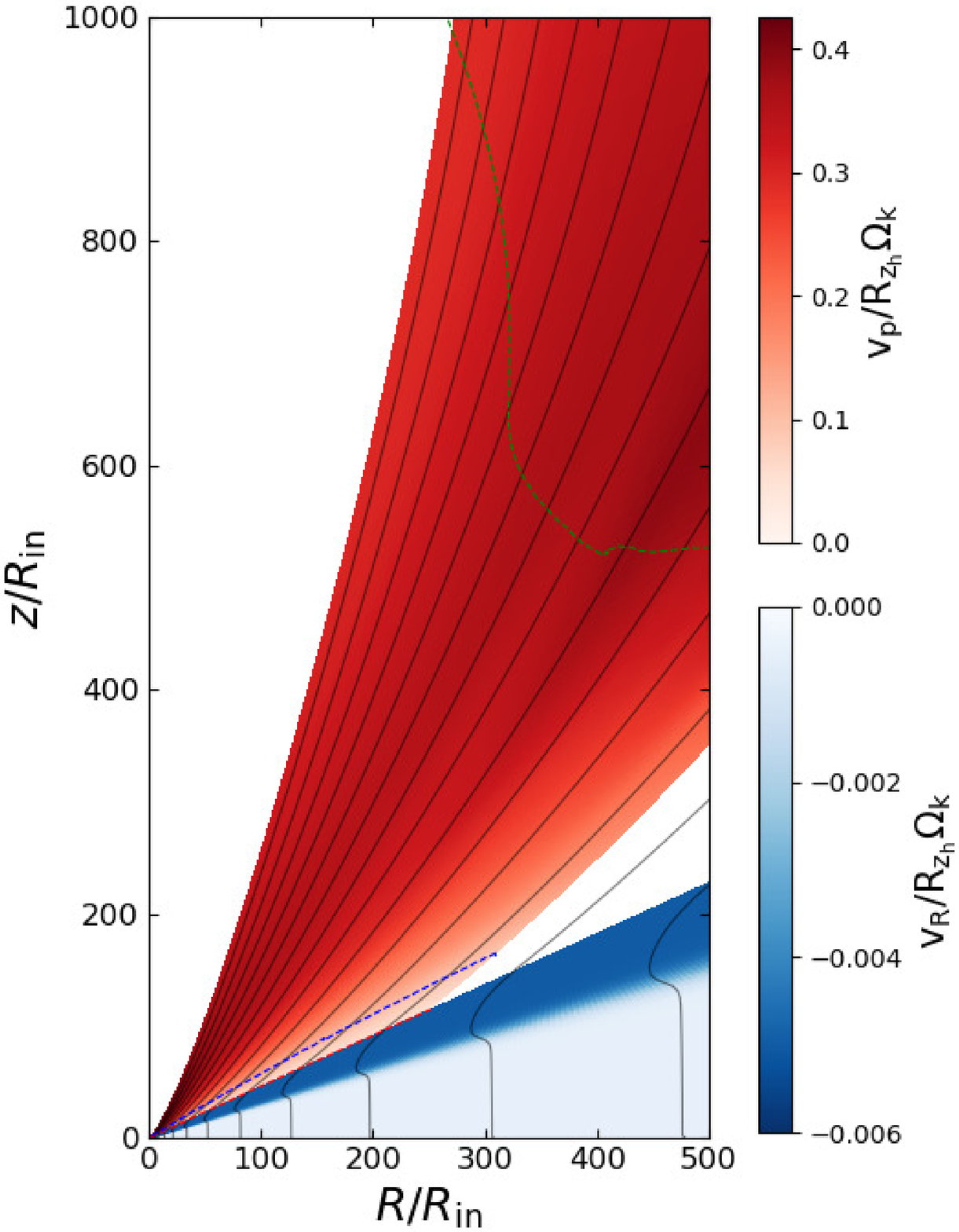}{0.4\textwidth}{}
    \fig{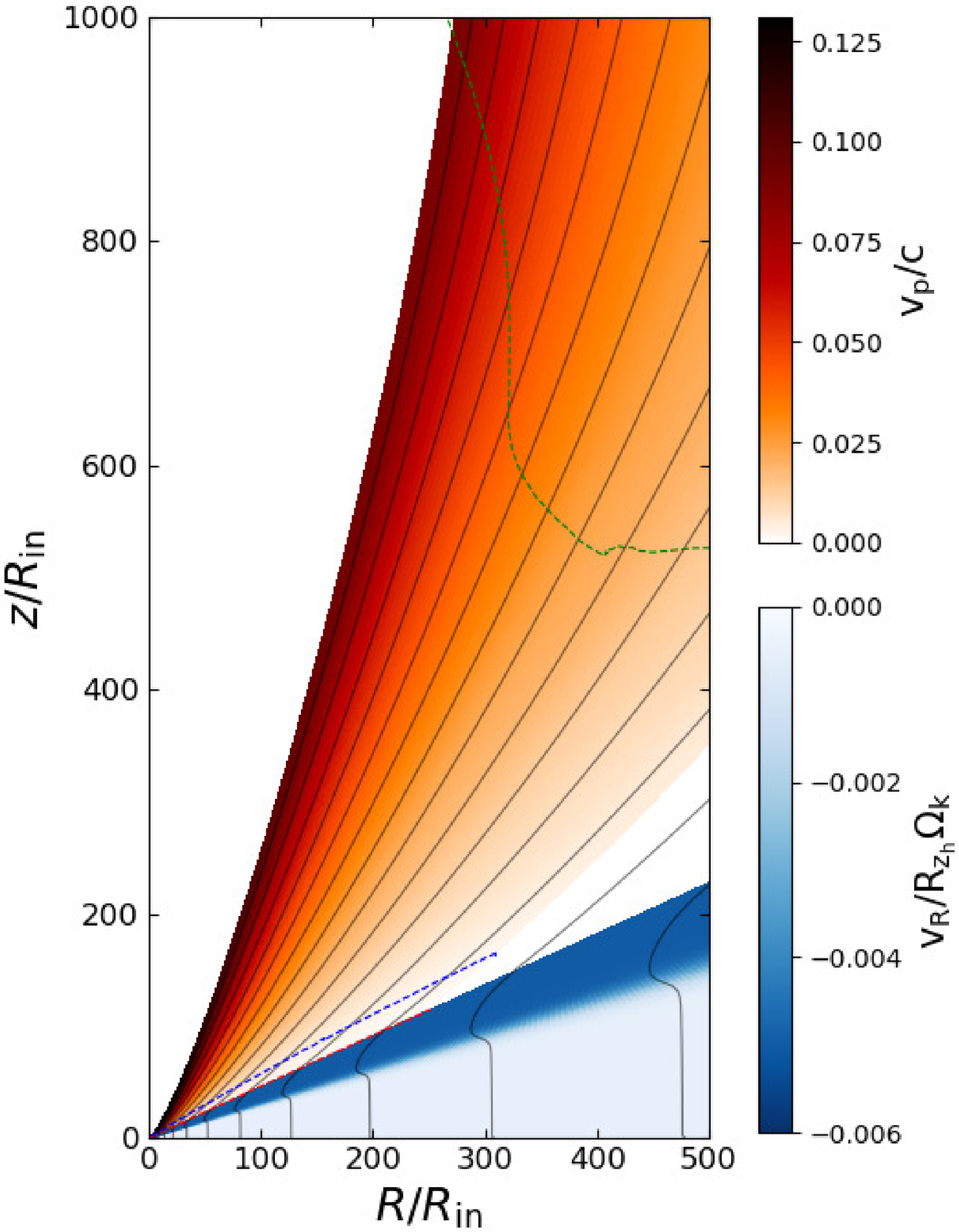}{0.4\textwidth}{}}
    
    \gridline{
    \fig{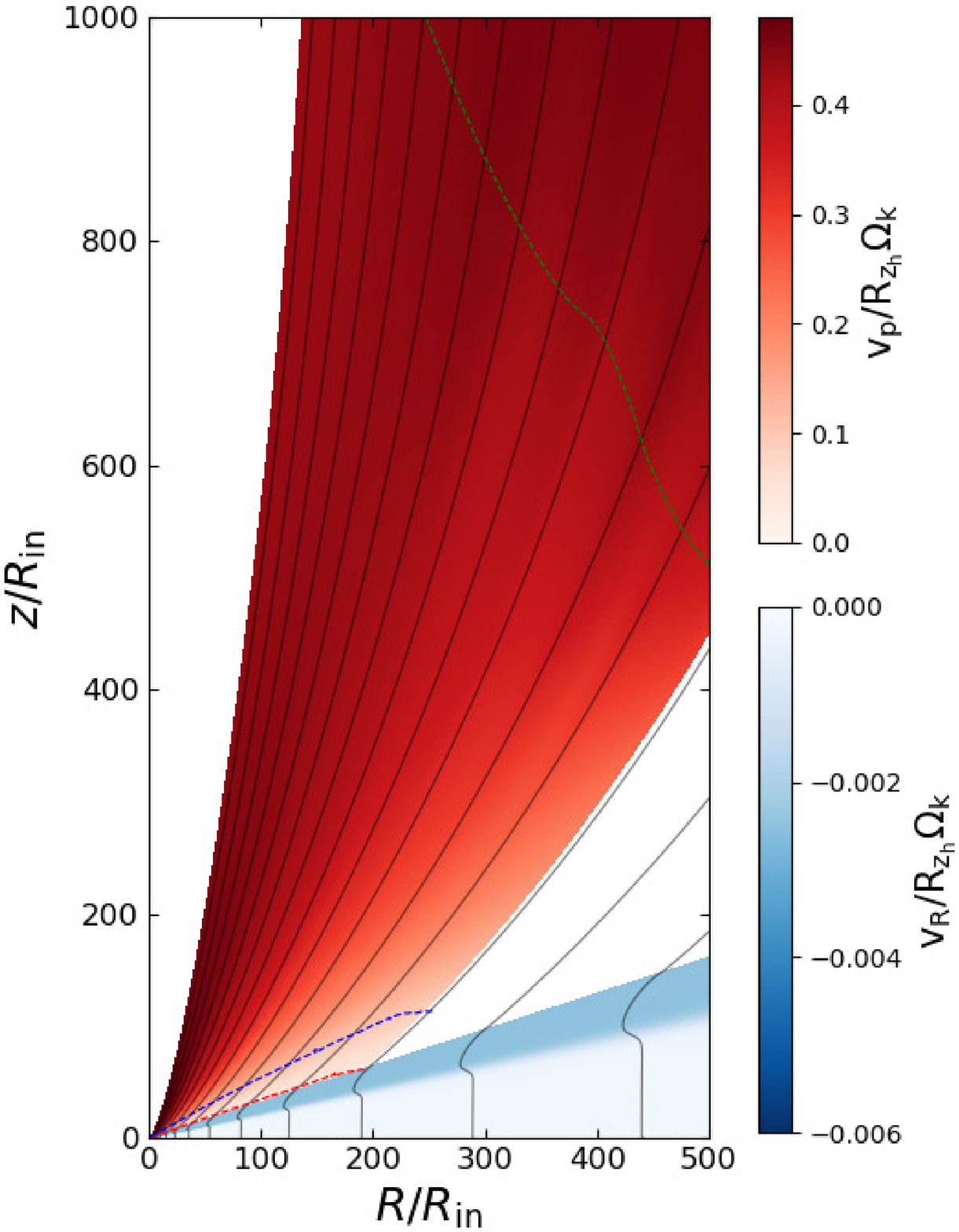}{0.4\textwidth}{}
    \fig{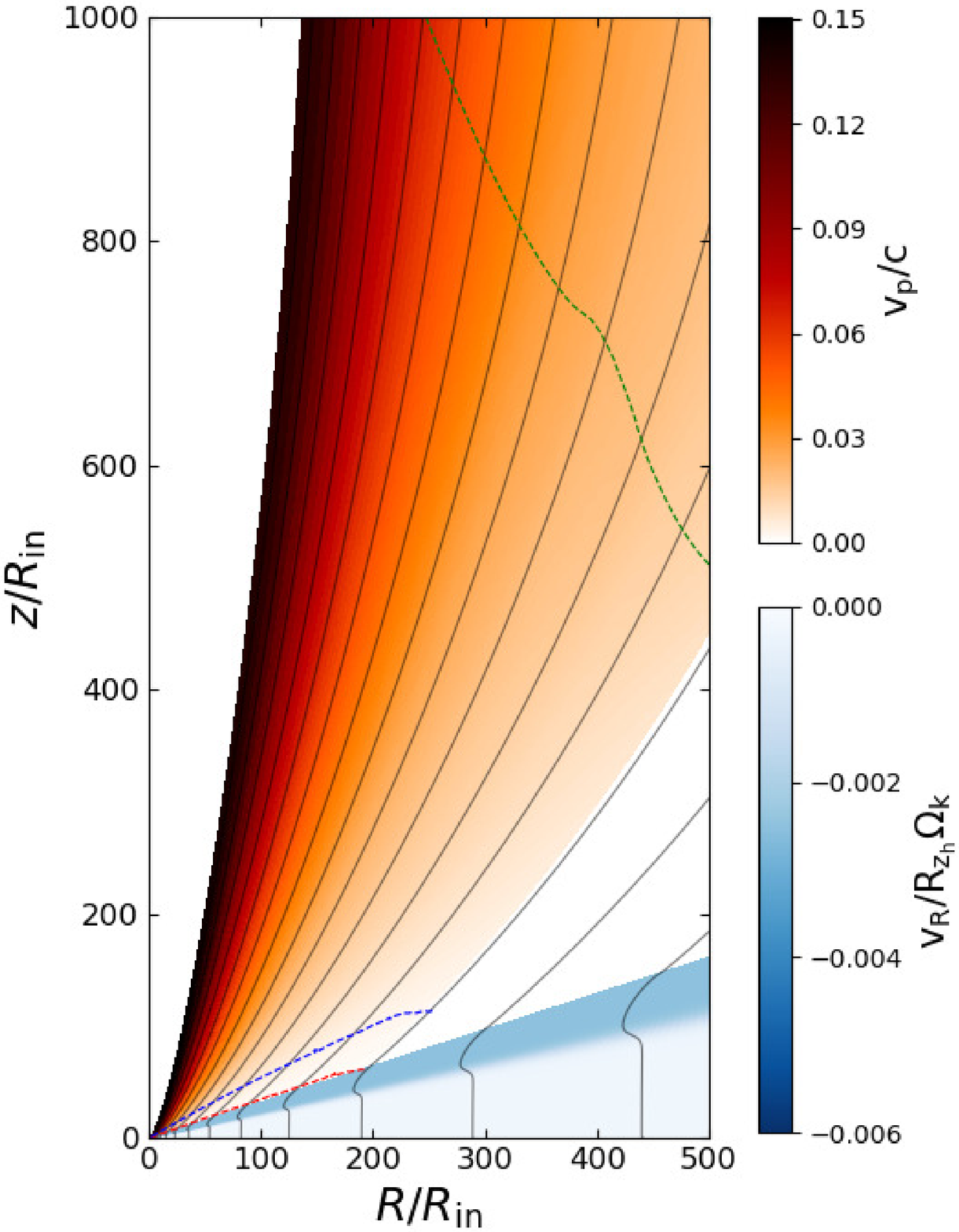}{0.4\textwidth}{}}
    
    \caption{The velocity profiles of the disks and the outflows.  The upper panels are calculated for case 1: $\Theta_{0}=0.005, \Theta_{z_{\rm h}}=0.05$, while the lower panels are calculated for case 2: $\Theta_{0}=0.0025, \Theta_{z_{\rm h}}=0.025$. The blue region show the radial velocities of the disks $v_{R}$, while the red areas show the derived poloidal velocities of the outflows $v_{\rm p}$ ($c$ is the light speed and $R_{z_{\rm h}} \Omega_{\rm k}$ is the local Keplerian velocity at the footpoint of the magnetic field line). The dashed lines correspond to the position of the slow sonic point (red), Alfv\'en point (blue) and the fast magnetosonic point (green), respectively. The solid curves indicate the poloidal magnetic field lines.}
    \label{fig:vp}
\end{figure}

The radial velocities of the disk-corona and the poloidal velocities of the outflows are plotted in Figure \ref{fig:vp}. The poloidal velocity $v_{\rm p}$, parallel to the poloidal field line (solid curves), of the outflow is scaled by the local Keplerian velocity of the field line foot point $R_{z_{\rm h}}\Omega_{\rm k}$ and the light speed $c$, respectively. The fast inward moving corona is responsible for the radial magnetic flux transport. We assume that the outflows are launched from the surface of the corona (i.e., $z=z_{\rm h}$),  which serves as a reservoir of the gas to feed the outflows. In this case, the magnetic flux is mainly dragged inward by the corona, thus the gas-to-magnetic ratio at the corona surface should be greater than unity, which means the outflows are driven by a relatively weak magnetic field (compared with gas pressure) \citep[][]{2018MNRAS.473.4268C}. The derived slow sonic points of the outflows are therefore very close to the surface of the corona (see red-dashed line in Figure \ref{fig:vp}).

The outflows accelerated by the magnetic field pass through the fast magnetosonic surface (see the green dashed lines in Figure \ref{fig:vp}), the terminal poloidal velocities of the outflows can be as high as $\sim 0.01c-0.1c$, which may account for some fast outflows observed in the XRBs and AGNs. The terminal speeds of the outflows for case 2: $\Theta_{0}=0.0025,\Theta_{z_{\rm h}}=0.025$ is slightly larger than that of the case 1: $\Theta_{z_{\rm h}}=0.005, \Theta_{z_{\rm h}}=0.05$ (Figure \ref{fig:vp}, see also the solid lines in the left panel of Figure \ref{fig:vpvphi_macc}). This is caused by the fact that the field line inclination angle of case 2 is slightly larger than that of case 1 and the resultant a deeper effective potential barrier along the field line, and correspondingly a smaller mass loss rate [$\dot{M}_{\rm w}= \dot{M}_{\rm out}-\dot{M}_{\rm acc}=(1-\dot{M}_{\rm acc}/\dot{M}_{\rm out})\times \dot{M}_{\rm out}$].

\begin{figure}
    \gridline{\fig{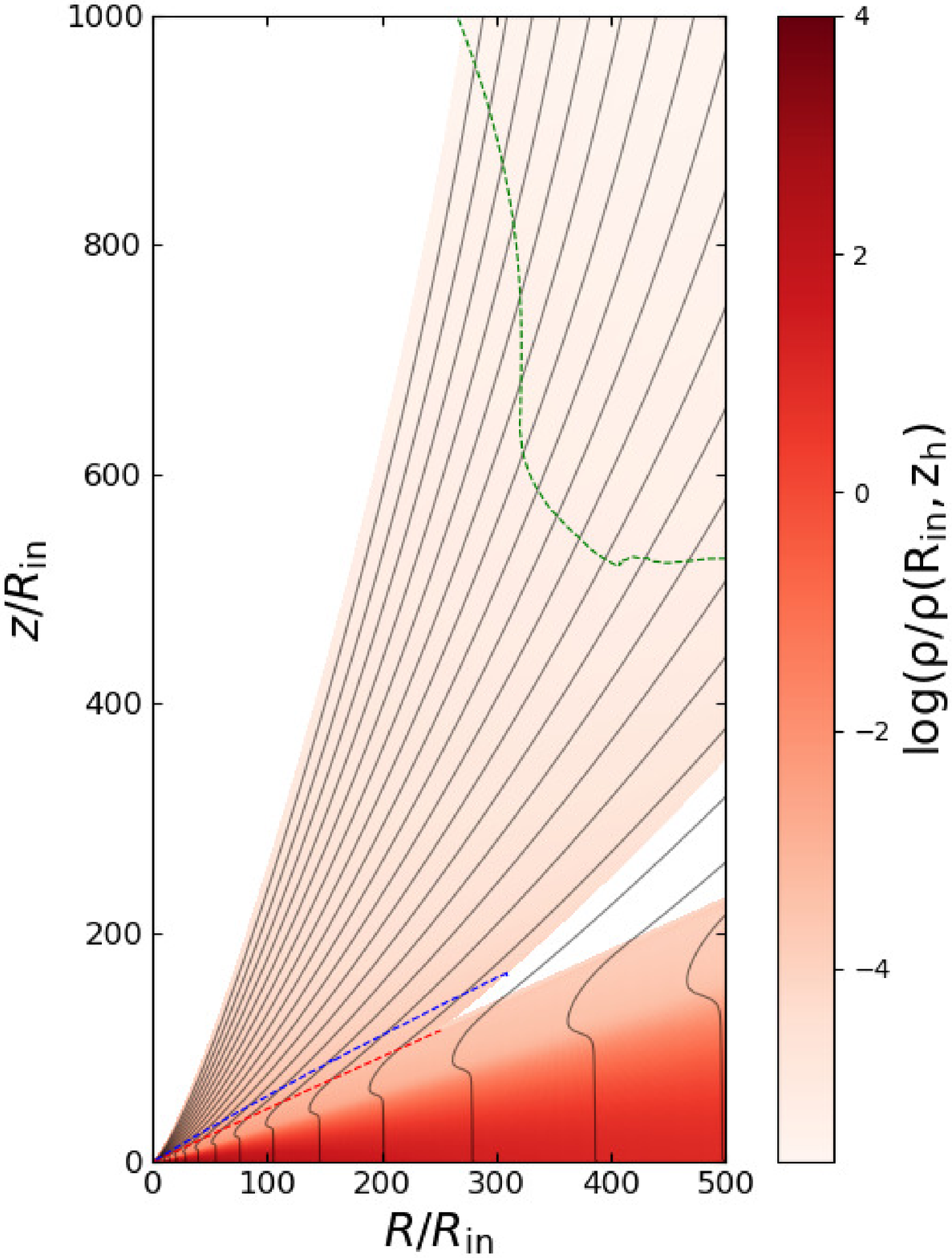}{0.4\textwidth}{}
    \fig{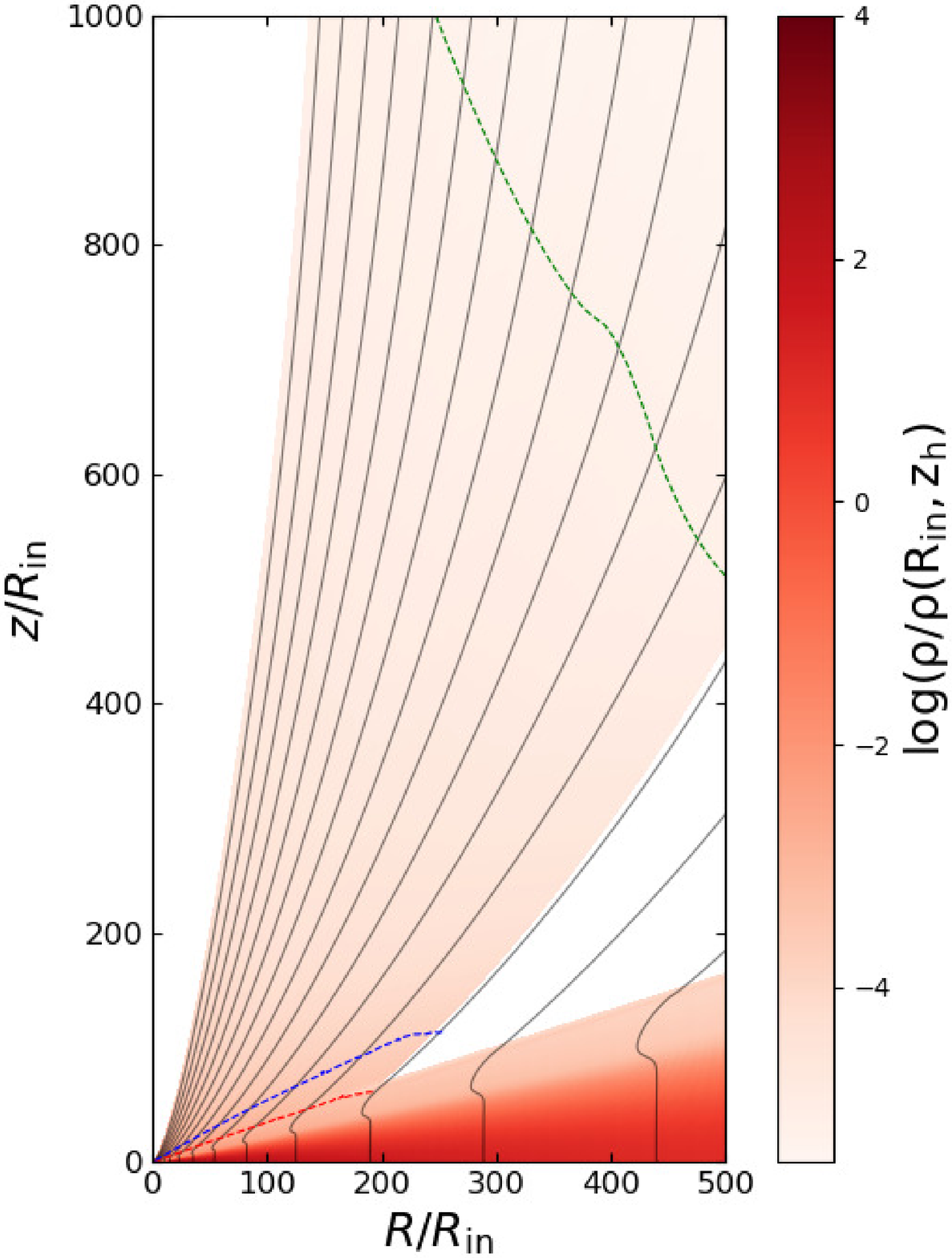}{0.4\textwidth}{}}
    \caption{The same as Figure \ref{fig:B_line} but for the density profiles of the disks and the outflows, and  $\rho(R_{{\rm in}},z_{\rm h})$ is the gas densities of the gas at the upper corona surface (i.e., $z=z_{\rm h}$) at radius $R=R_{\rm in}$.}
    \label{fig:Den}
\end{figure}

\begin{figure}
    \gridline{\fig{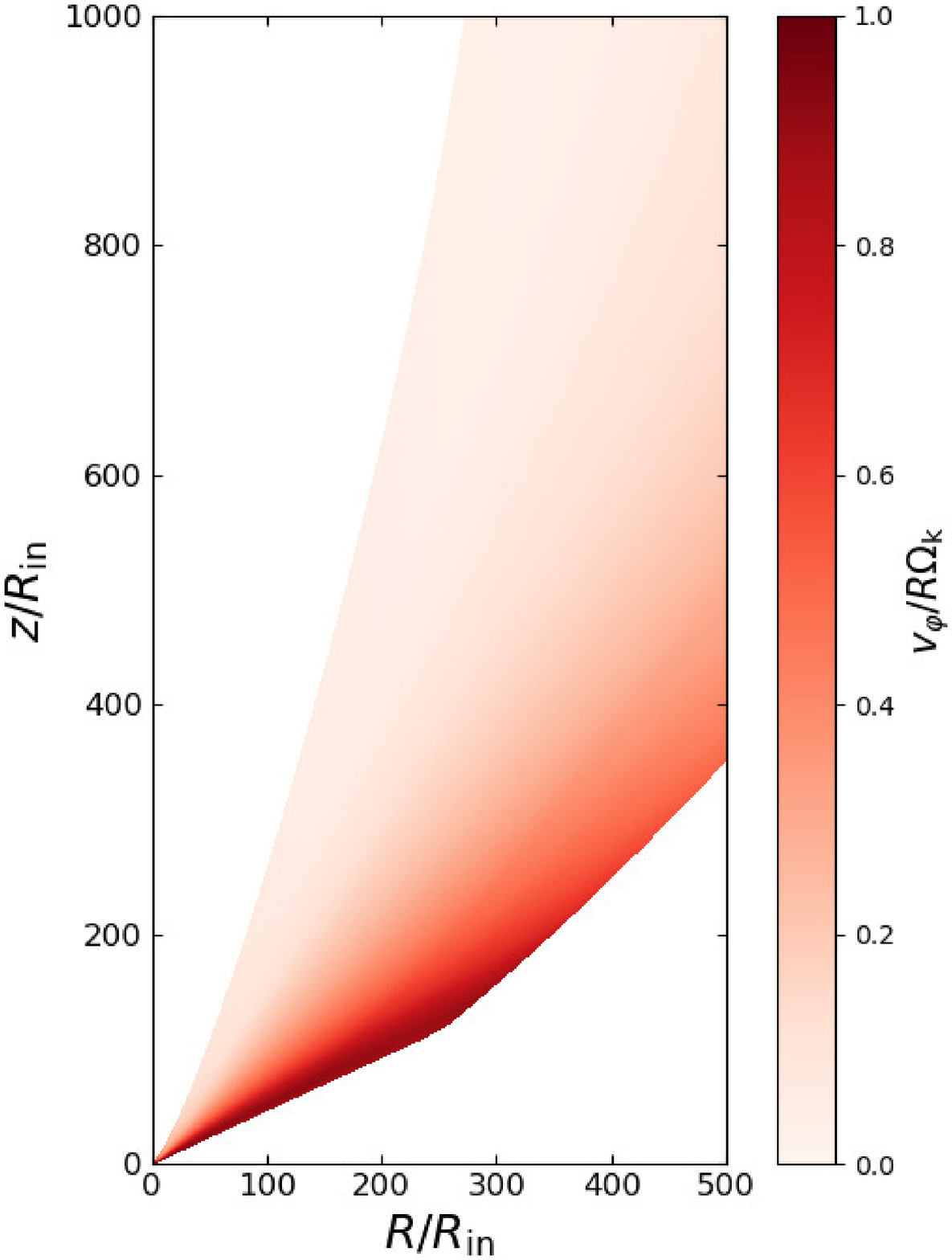}{0.4\textwidth}{}
    \fig{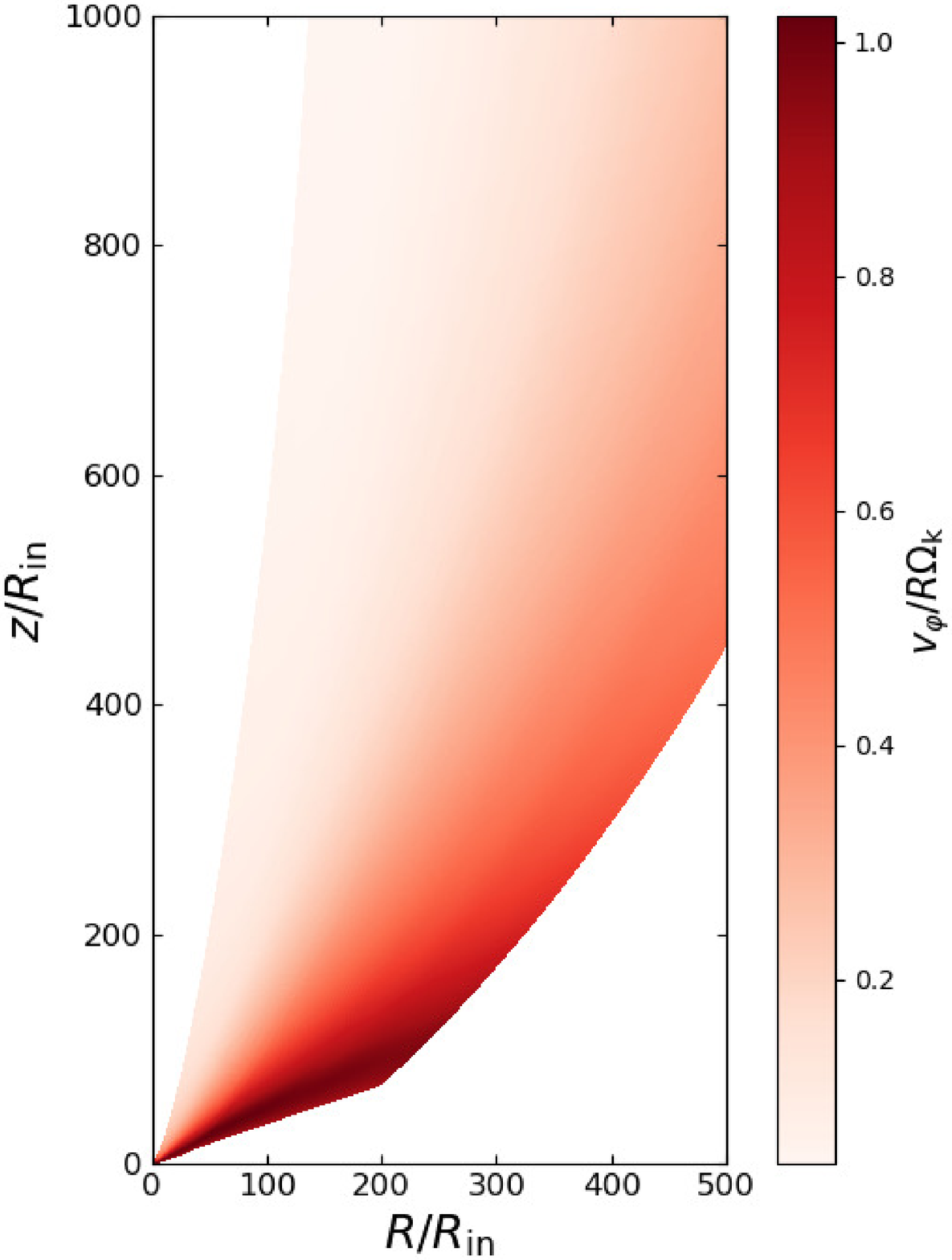}{0.4\textwidth}{}}
    \caption{The same as Figure \ref{fig:B_line} but for the toroidal velocity profiles of the outflows.}
    \label{fig:vphi}
\end{figure}

The densities of the disks and outflows are plotted in Figure \ref{fig:Den}. The central high density cold disks are covered with low density hot coronas, which feed a fraction of the gas into the outflows. The toroidal velocities are plotted in Figures \ref{fig:vphi}. In the region close to Alfv\'en surface the toroidal velocity of the outflow reach the maximum value, beyond the Alfv\'en surface the toroidal velocity decrease very fast due to the field line being unable to enforce the gas co-rotating with the field anymore.

\begin{figure}
    \gridline{\fig{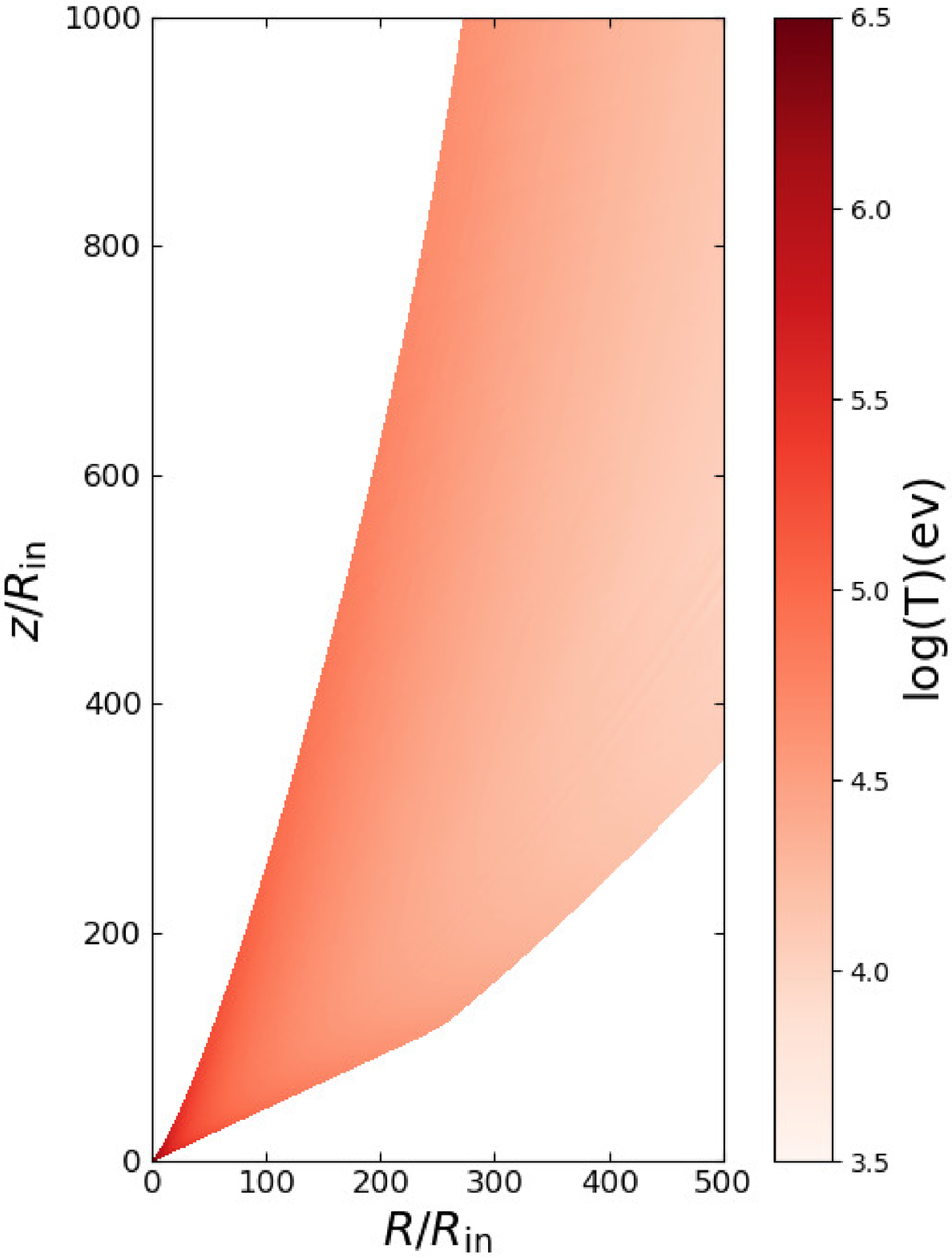}{0.4\textwidth}{}
    \fig{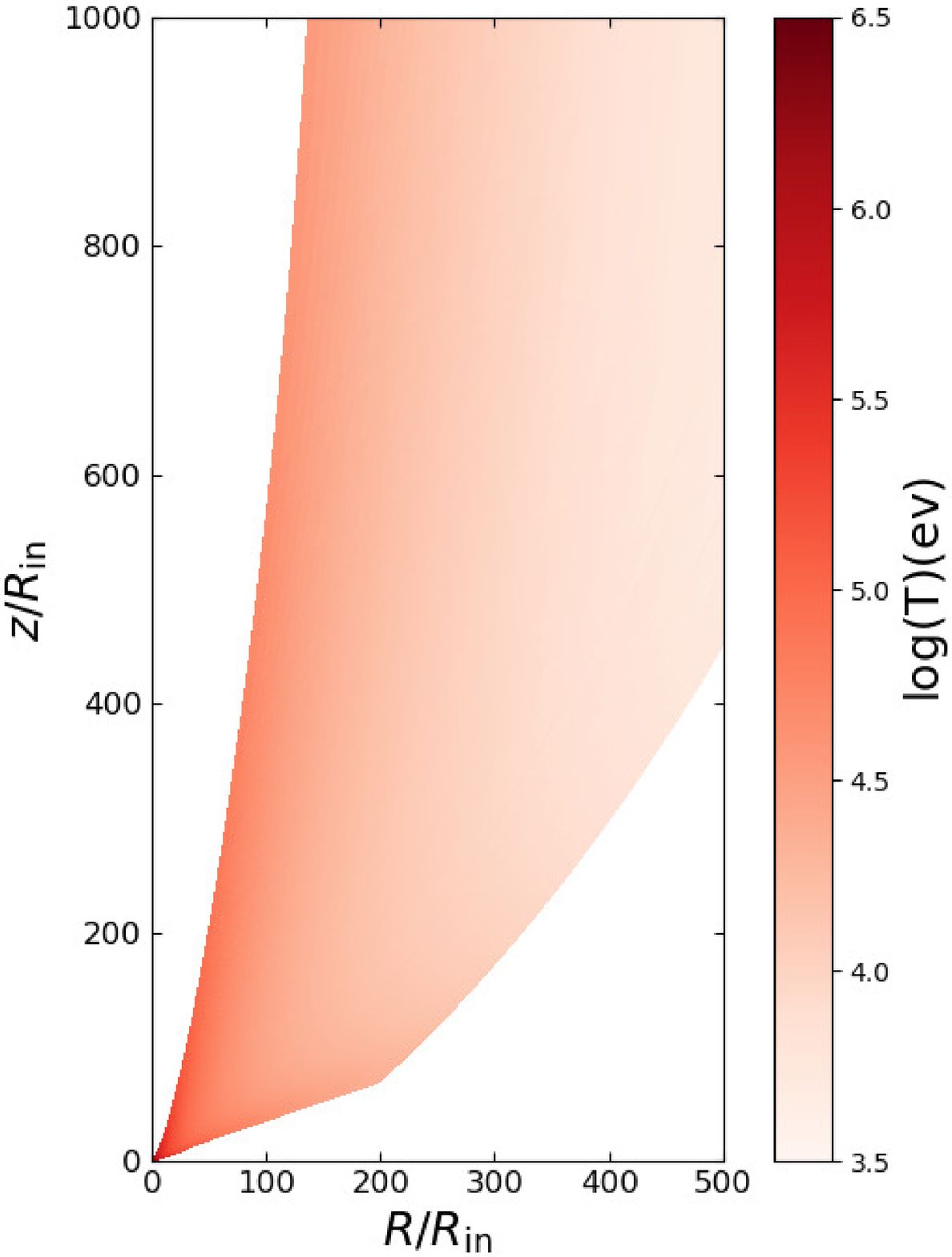}{0.4\textwidth}{}}
    \caption{The same as Figure \ref{fig:B_line} but for the gas temperatures (in units of electron-volt) of the outflows.}
    \label{fig:Tem}
\end{figure}

The temperatures of the outflows are plotted in figure \ref{fig:Tem}, which is calculated with a polytropic Equation of state, and we adopt an adiabatic index $\gamma=1.4$ in most of our calculations as that adopted in the numerical simulation work done by \cite{2018ApJ...857...34Z}. The temperatures of the outflows drop quickly (about several orders of magnitude) along the filed line (see the left panel of Figure \ref{fig:Tem} and right panel of Figure \ref{fig:vphi1_den1_tem1}), but it can still reach several ten keV at a large distance from the disk (see Figure \ref{fig:Tem}), which means that the gas in the outflows accelerated from the surface of the corona may naturally reach a high ionization state. However, we should be cautious, because the temperature of the outflows depends on the cooling and heating processes in the outflows, which have not been considered in this work. We also give the results calculated with $\gamma=5/3$ for comparison, and find that the results are not sensitive to the value of $\gamma$. The slow sonic points and the Alfv\'en points are almost remain unchanged see the left panel of Figures \ref{fig:Den}, \ref{fig:vphi1_den1_tem1} and \ref{fig:vpvhi1_dmacc1}. The accretion rate of the disk-corona does not show a discernible variation (see right panel of Figure \ref{fig:vpvhi1_dmacc1}), while only slightly shifts to large radii for the fast sonic points of the outflows. 

\begin{figure}
    \gridline{
    \fig{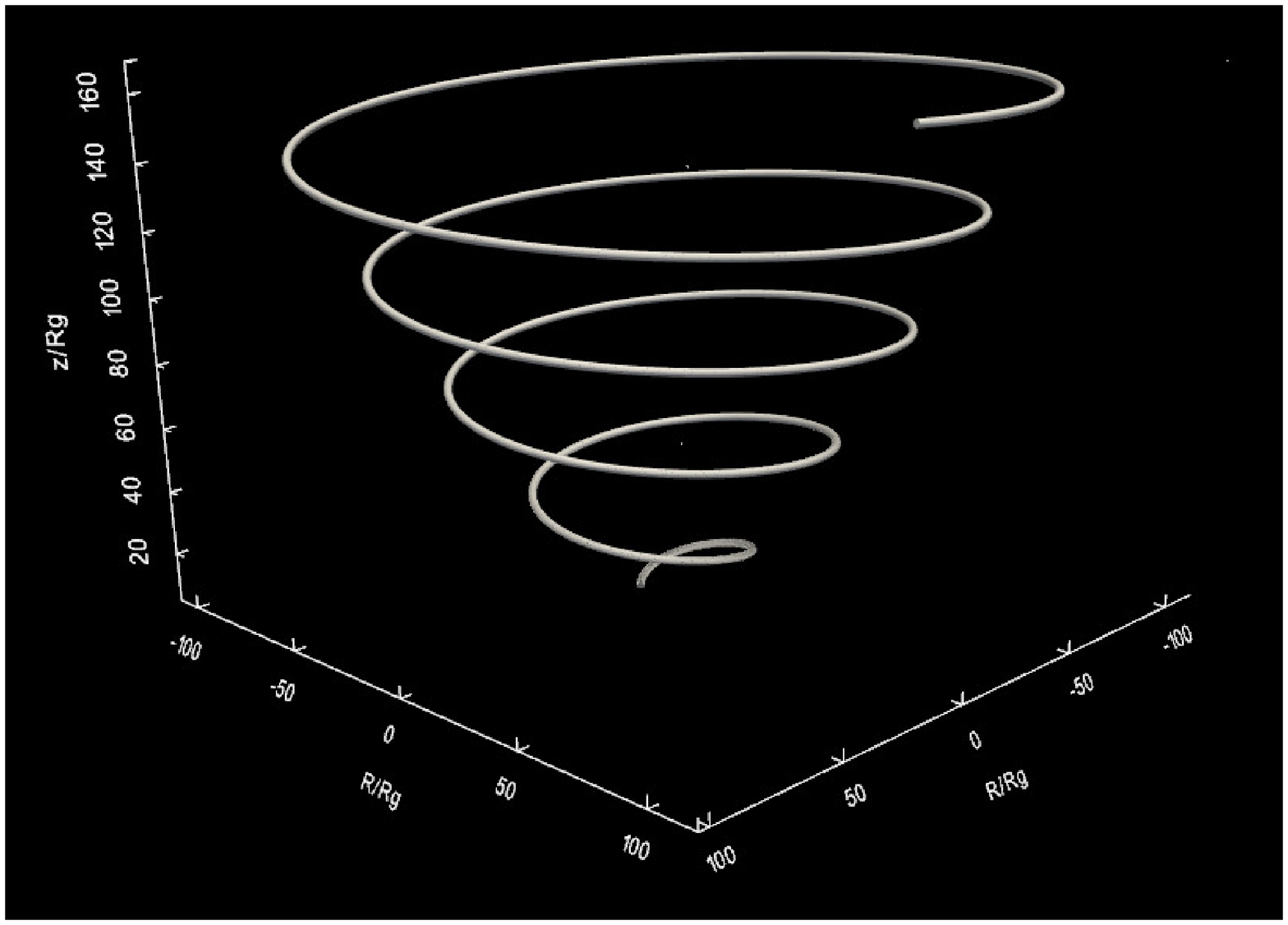}{0.45\textwidth}{}
    \fig{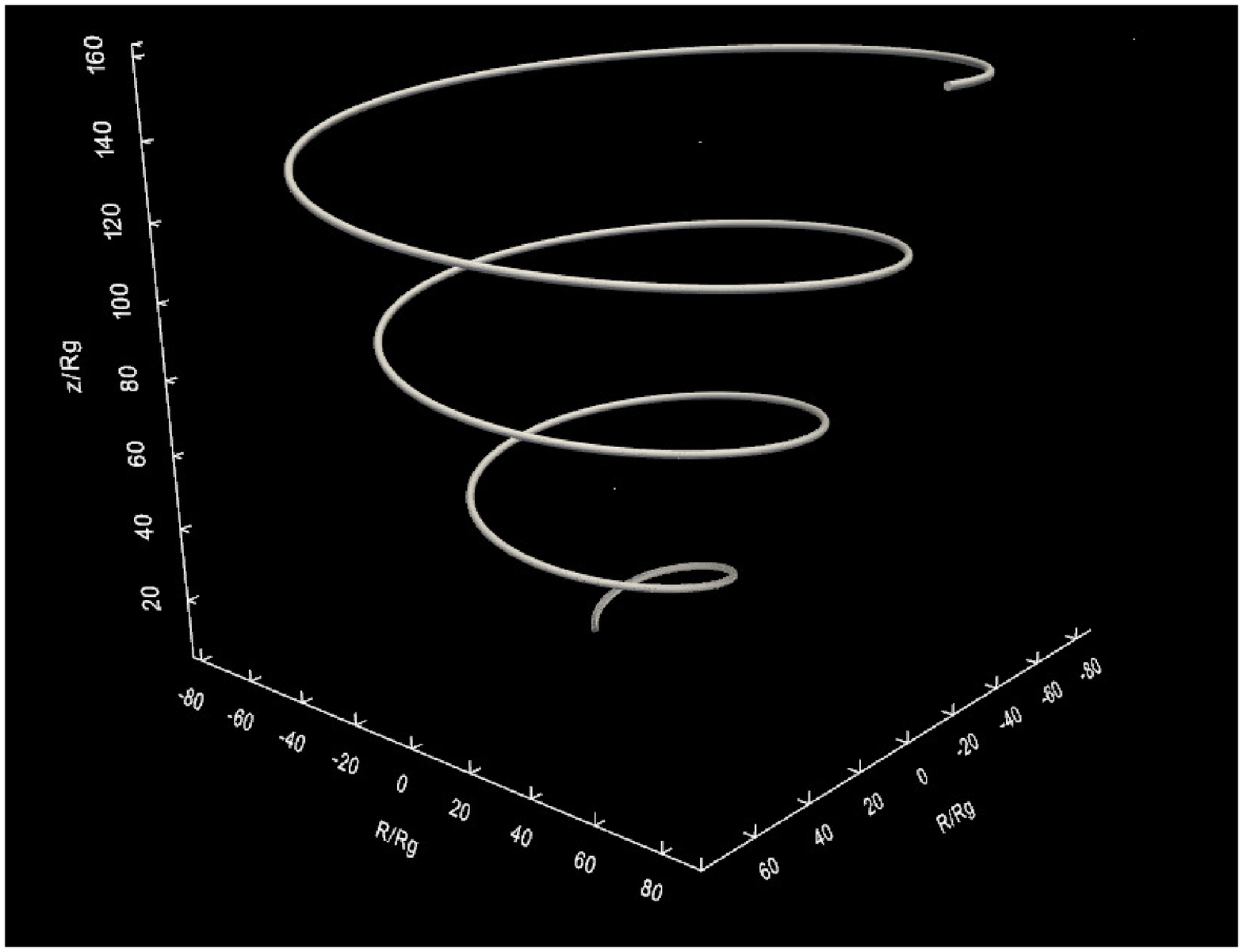}{0.43\textwidth}{}}
    
    \gridline{
    \fig{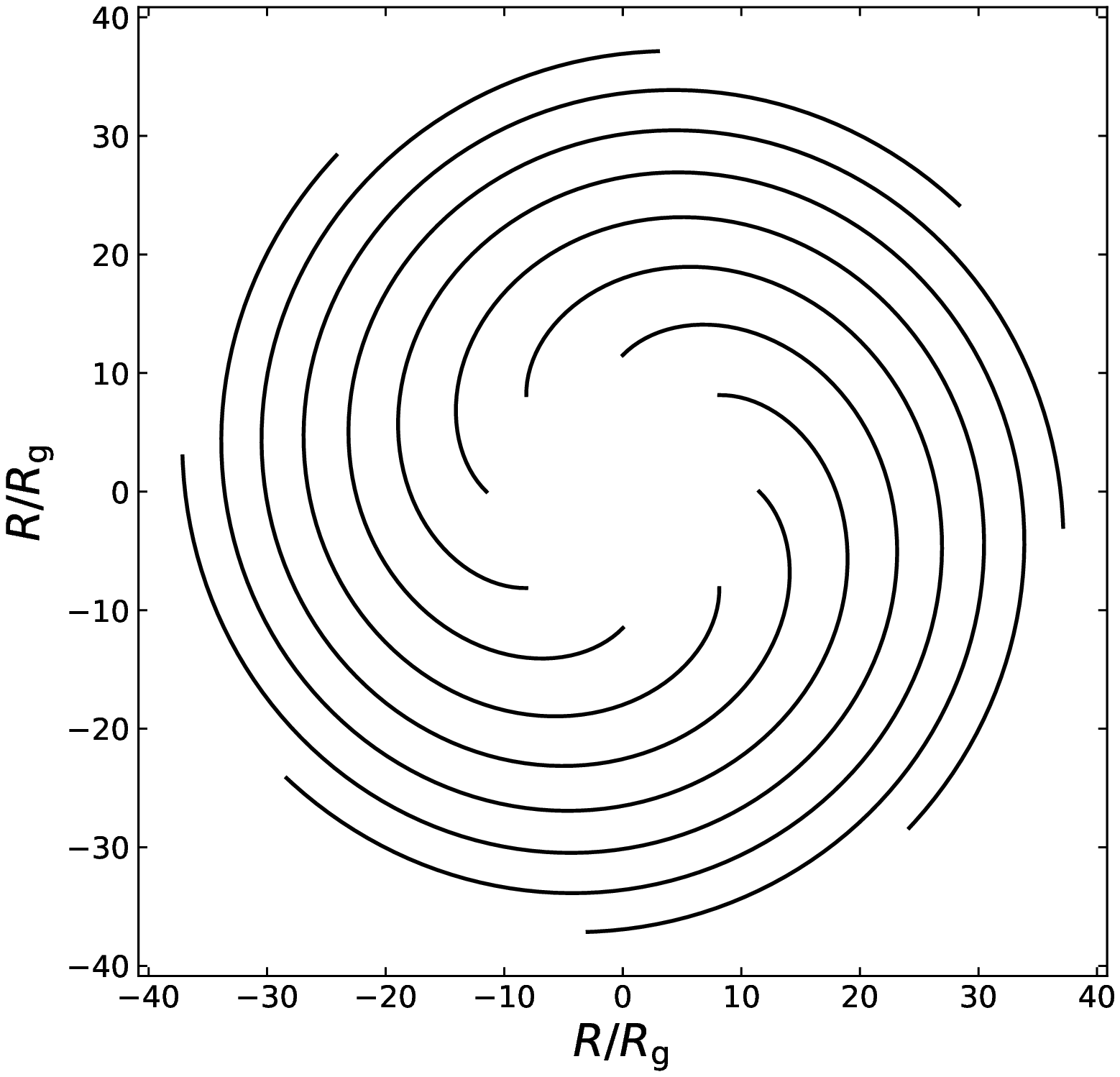}{0.45\textwidth}{}
    \fig{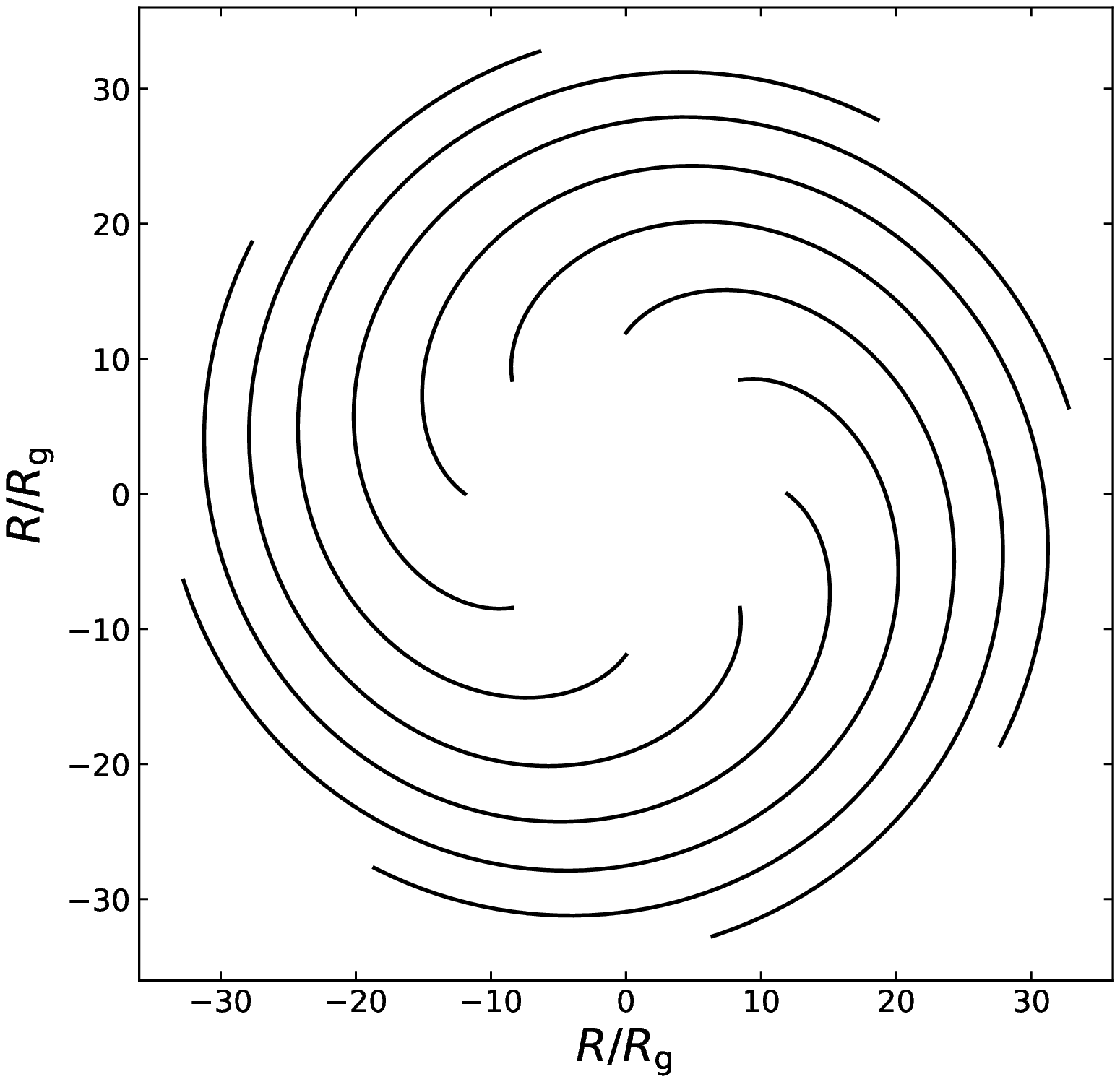}{0.45\textwidth}{}}
    
    \caption{The same as Figure \ref{fig:B_line}, but for the configurations of the field lines. Upper panels are the spatial structure of the field lines, and the radius of the field line foot point is about $R/R_{\rm in} \sim 2$. Lower panels show the projection of the field line onto the disk mid-plane. Two adjacent lines have a toroidal angle difference of $\pi/4$.}
    \label{fig:2D3Dline}
\end{figure}

\begin{figure}
    \gridline{\fig{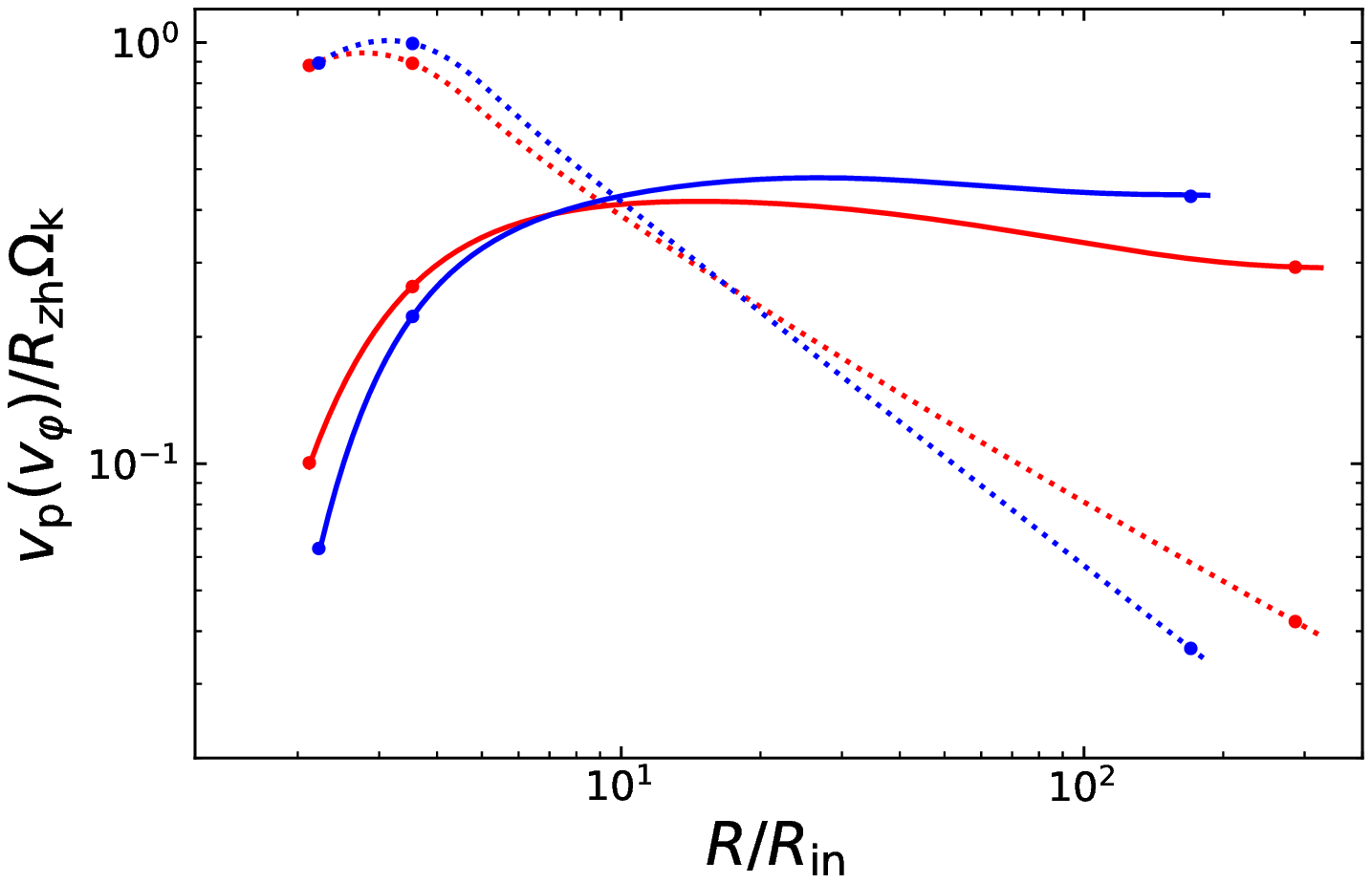}{0.45\textwidth}{}
    \fig{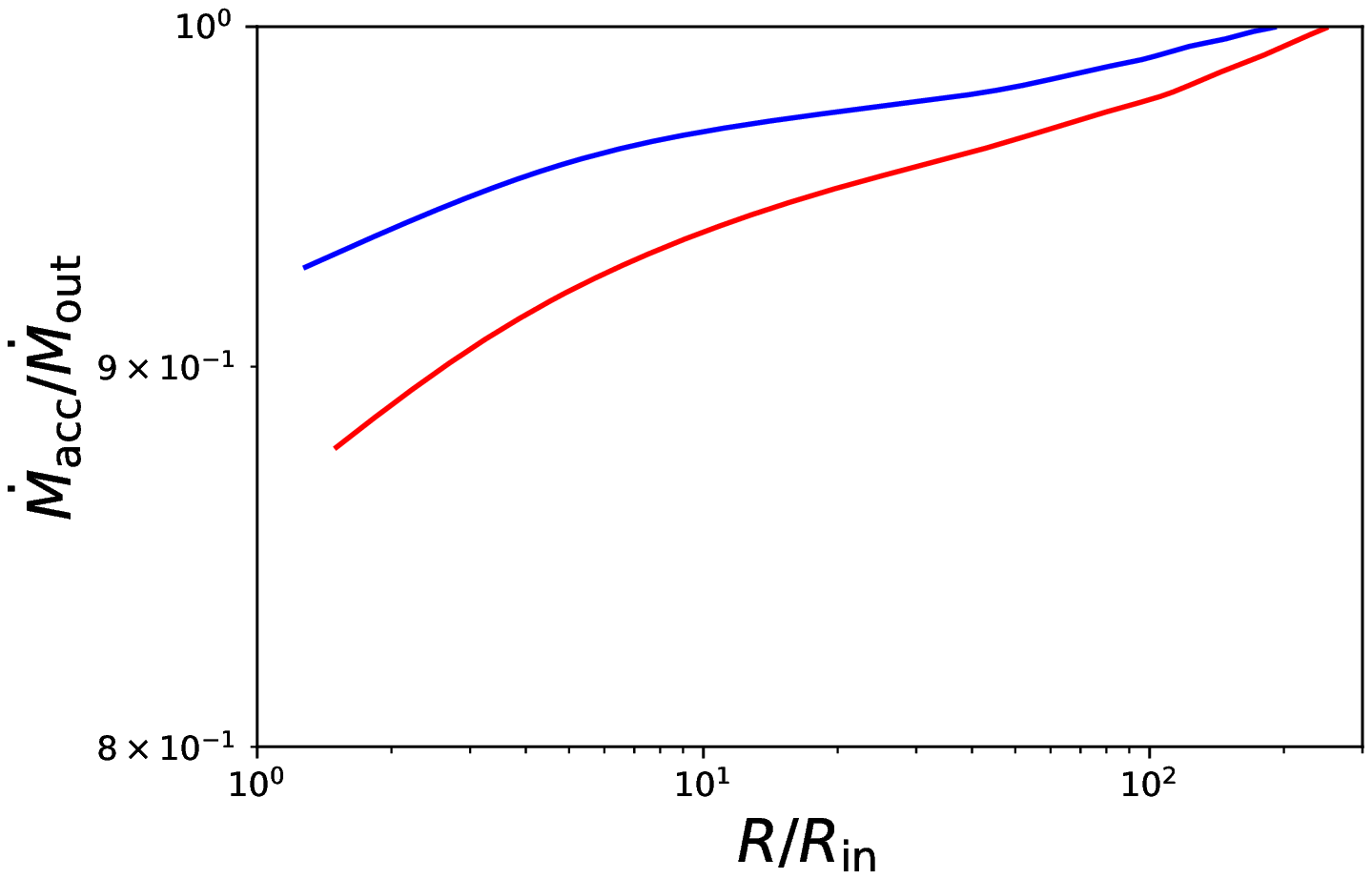}{0.45\textwidth}{}}
    \caption{The same as Figure \ref{fig:Bz}. Left panel: the poloidal and toroidal velocity of the outflows along the magnetic field lines shown in the upper panels of Figure \ref{fig:2D3Dline}. The red lines correspond to the field line in upper-left panel of Figure \ref{fig:2D3Dline}, while the blue lines corresponds to the field line in upper-right panel of Figure \ref{fig:2D3Dline}. The solid lines indicate the poloidal velocities $v_{\rm p}$ of the outflows, while the dotted lines are the toroidal velocities $v_{\varphi}$. The solid points, from left to right, on each curves are the slow, Alfv\'en and fast points, respectively. Right panel: the mass accretion rates of the disks as functions of radius, where $\dot{M}_{\rm out}$ is the accretion rate at radius $R=R_{\rm out}$. The adiabatic index $\gamma=1.4$ is adopted in the calculations.}
    \label{fig:vpvphi_macc}
\end{figure}

The spatial structures of the field lines and its projection onto the equatorial plane are plotted in Figure \ref{fig:2D3Dline}. The emergent spectra (especially their polarization) of the disk-corona systems can be calculated when the field configuration/strength derived here is properly included. The poloidal and toroidal velocities of the outflows along a single field line (corresponding to the field line plotted in the upper panel of Figure \ref{fig:2D3Dline}) are plotted in the left panel of Figure \ref{fig:vpvphi_macc}.
In the right panel of Figure \ref{fig:vpvphi_macc}, the mass accretion rates within the disk are plotted, which shows only a small fraction ($\sim 7\%-12\%$) of the accreting gas in the disk has been driven into the outflows.

%%%%%%Fig vp%%%%%%
%%%%%%%%%%%%%%%%%%

%%%%%%Fig den%%%%%%
%%%%%%%%%%%%%%%%%%

%%%%%%Fig vphi%%%%%%
%%%%%%%%%%%%%%%%%%

%%%%%%Fig Tem%%%%%%
%%%%%%%%%%%%%%%%%%

%%%%%%2D 3D line%%%%%%
%%%%%%%%%%%%%%%%%%

%%%%%%Fig vpvph_macc%%%%%%
%%%%%%%%%%%%%%%%%%

%%%%%%Fig vp1%%%%%%
%%%%%%%%%%%%%%%%%%
\begin{figure}
    \gridline{
    \fig{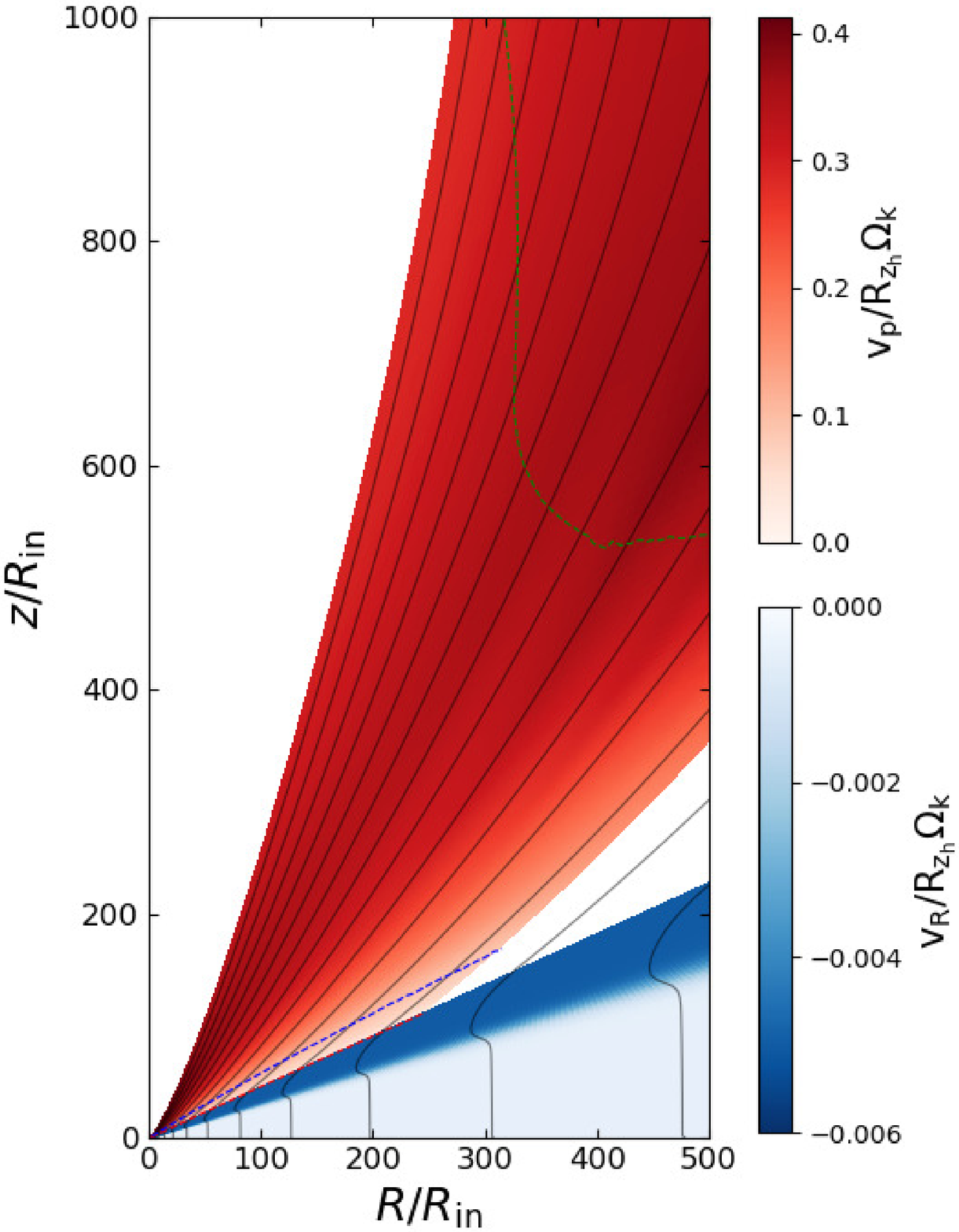}{0.4\textwidth}{}
    \fig{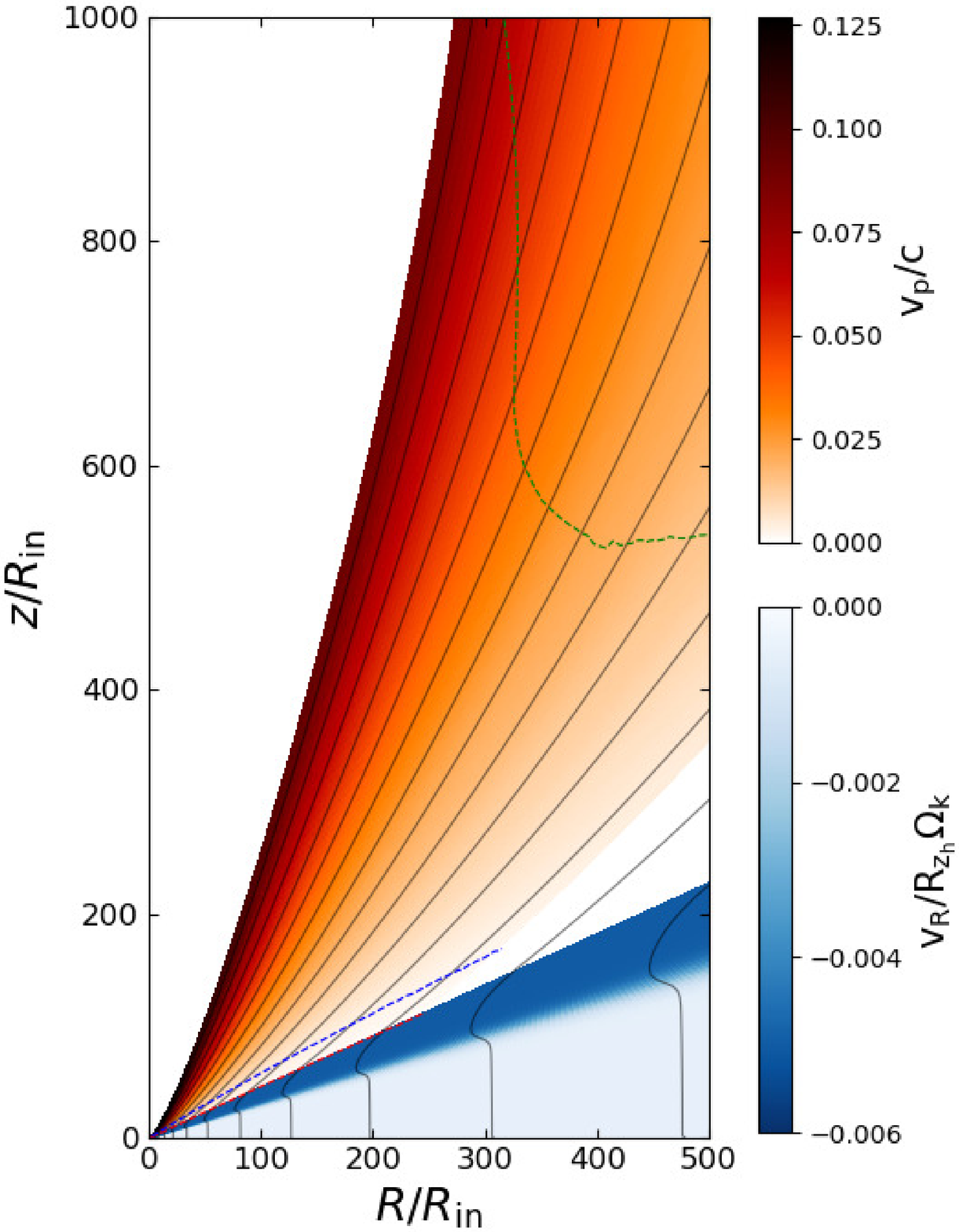}{0.4\textwidth}{}}
    \caption{Velocity profiles of the disks and the outflows. The parameters $\Theta_{0}=0.005\;\text{and}\;\Theta_{z_{\rm h}}=0.05$ are adopted, which is the same as the upper panel of Figure \ref{fig:vp} but $\gamma=5/3$ is adopted in the calculations.
    }
    \label{fig:vp1}
\end{figure}

%%%%%%vphi1_den1_tem1%%%%%%
%%%%%%%%%%%%%%%%%%
\begin{figure}
    \gridline{
    \fig{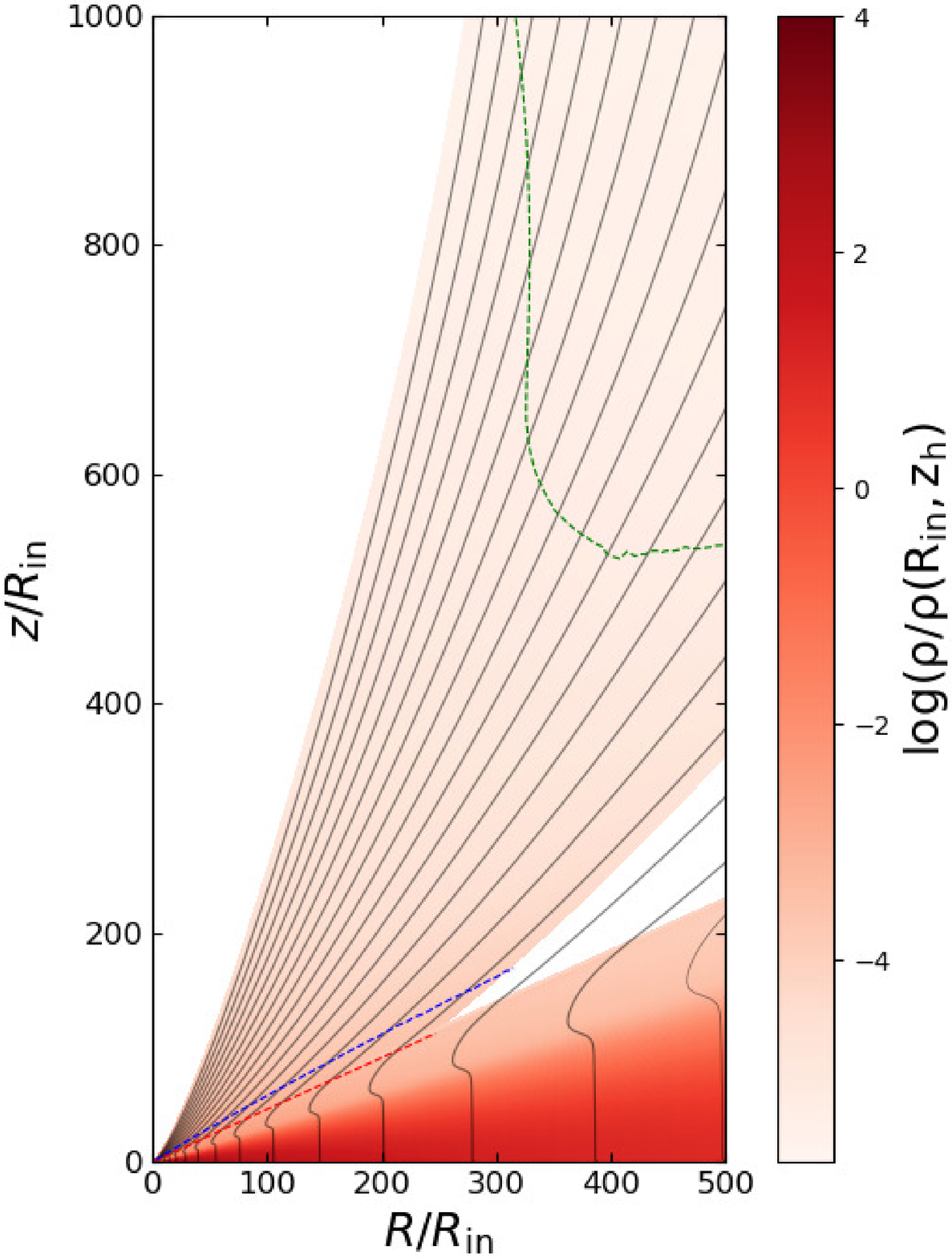}{0.33\textwidth}{}
    \fig{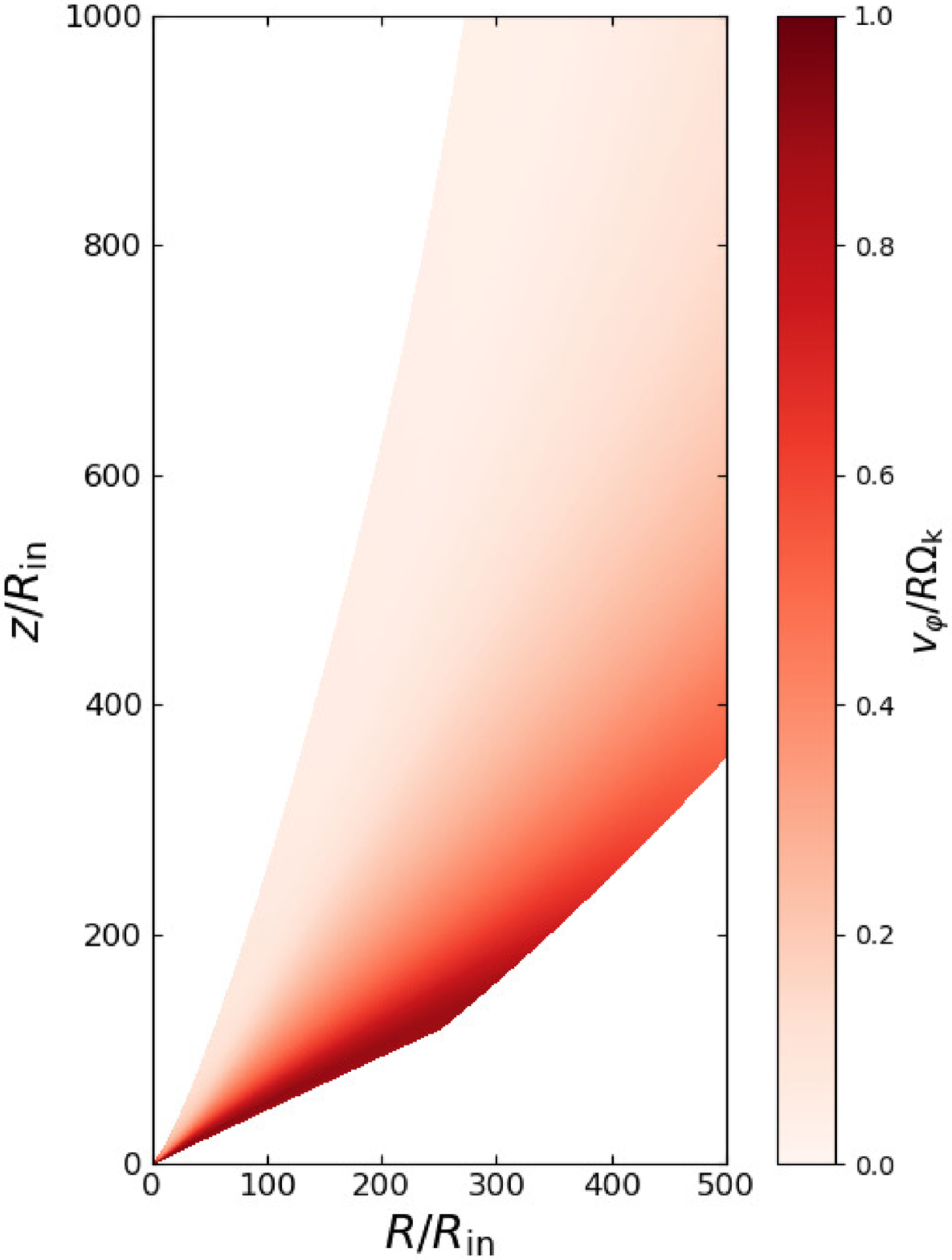}{0.33\textwidth}{}
    \fig{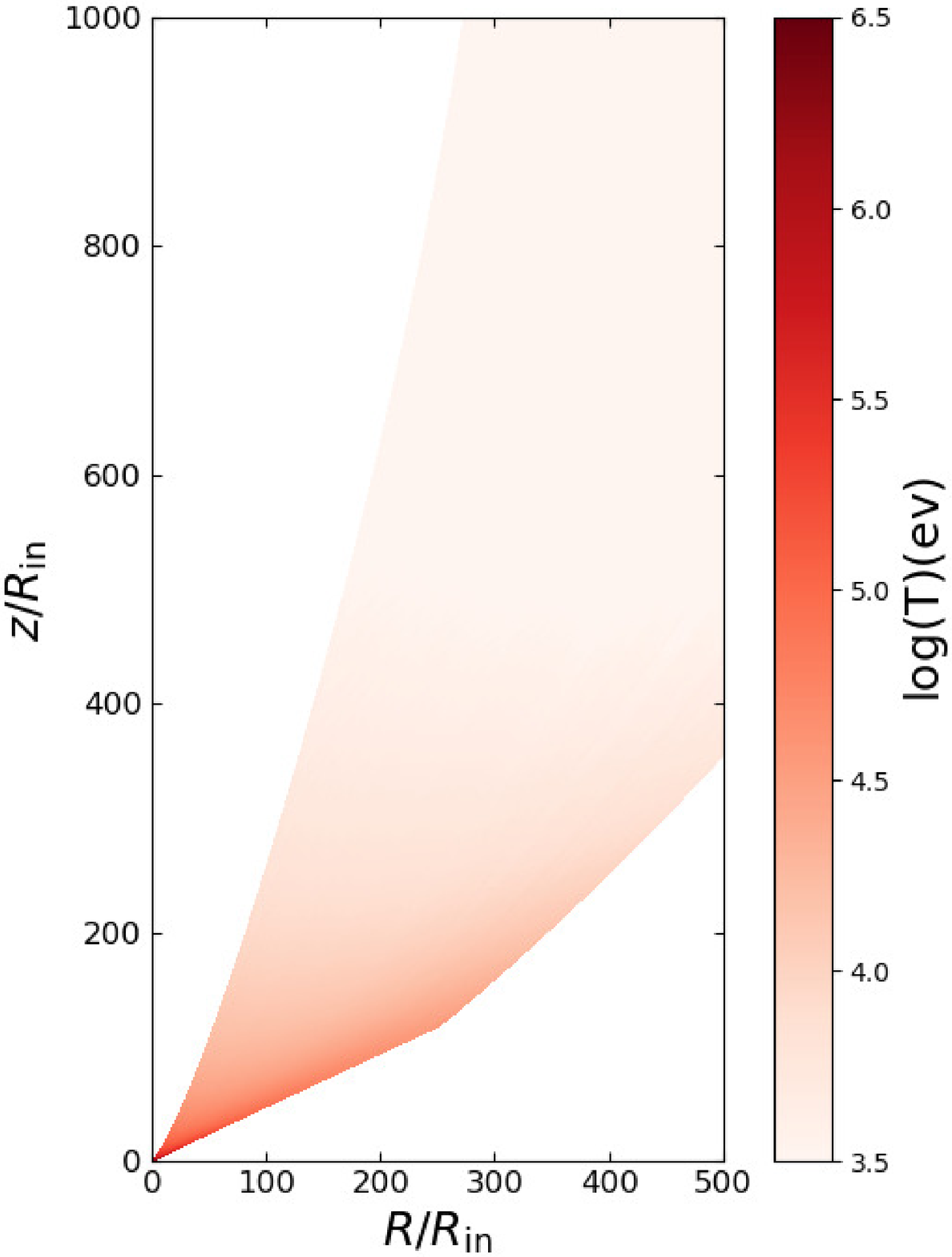}{0.33\textwidth}{}}
    \caption{Left panel: density profile, the same as the left panel of Figure \ref{fig:Den} but for $\gamma=5/3$. Central panel: toroidal velocity profile, the same as the left panel of Figure \ref{fig:vphi} but for $\gamma=5/3$. Right panel: temperature profile, the same as the left panel of Figure \ref{fig:Tem} but for $\gamma=5/3$.  }
    \label{fig:vphi1_den1_tem1}
\end{figure}

%%%%%%vpvhi1_dmacc1%%%%%%
%%%%%%%%%%%%%%%%%%
\begin{figure}
    \gridline{
    \fig{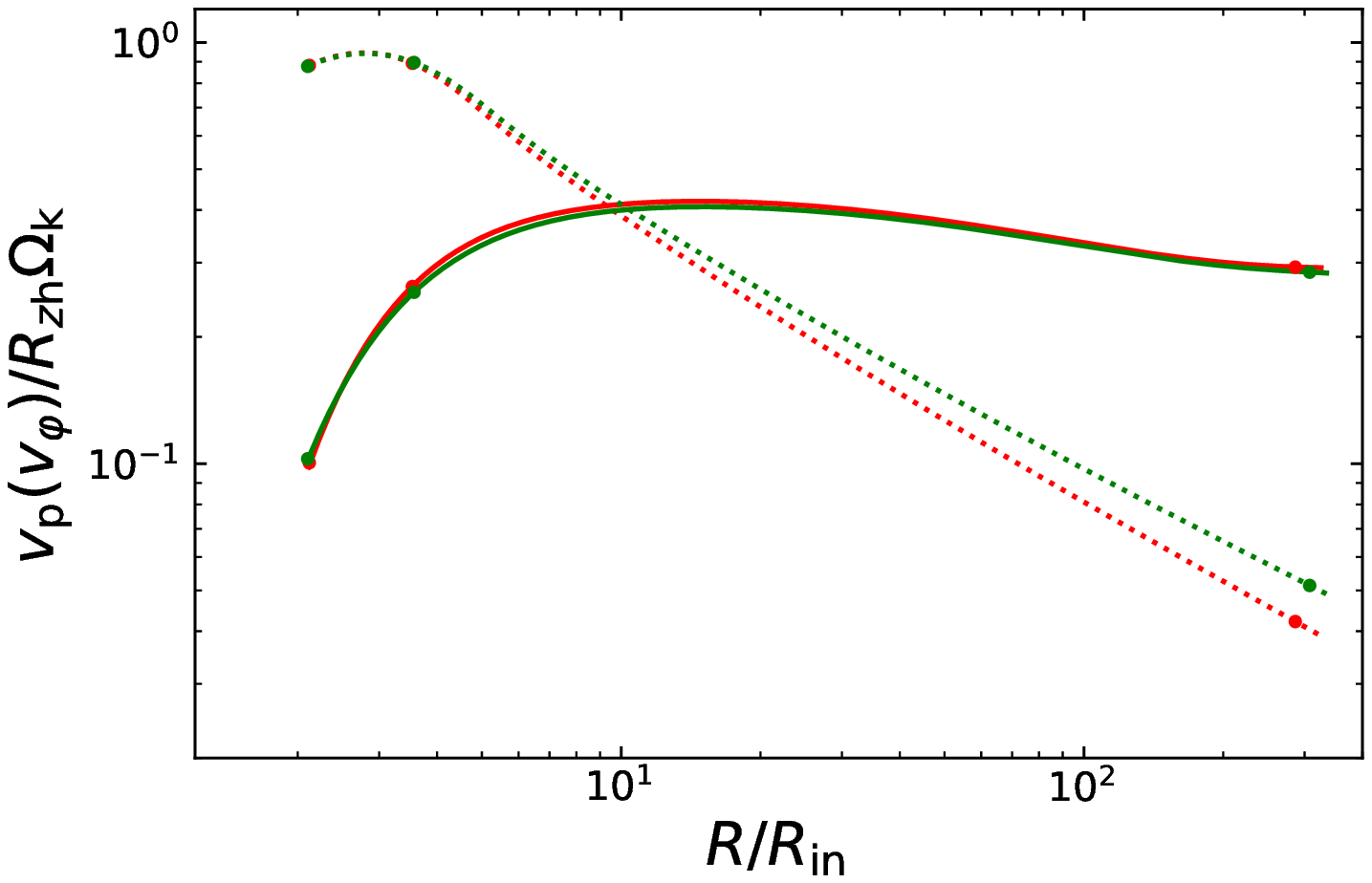}{0.5\textwidth}{}
    \fig{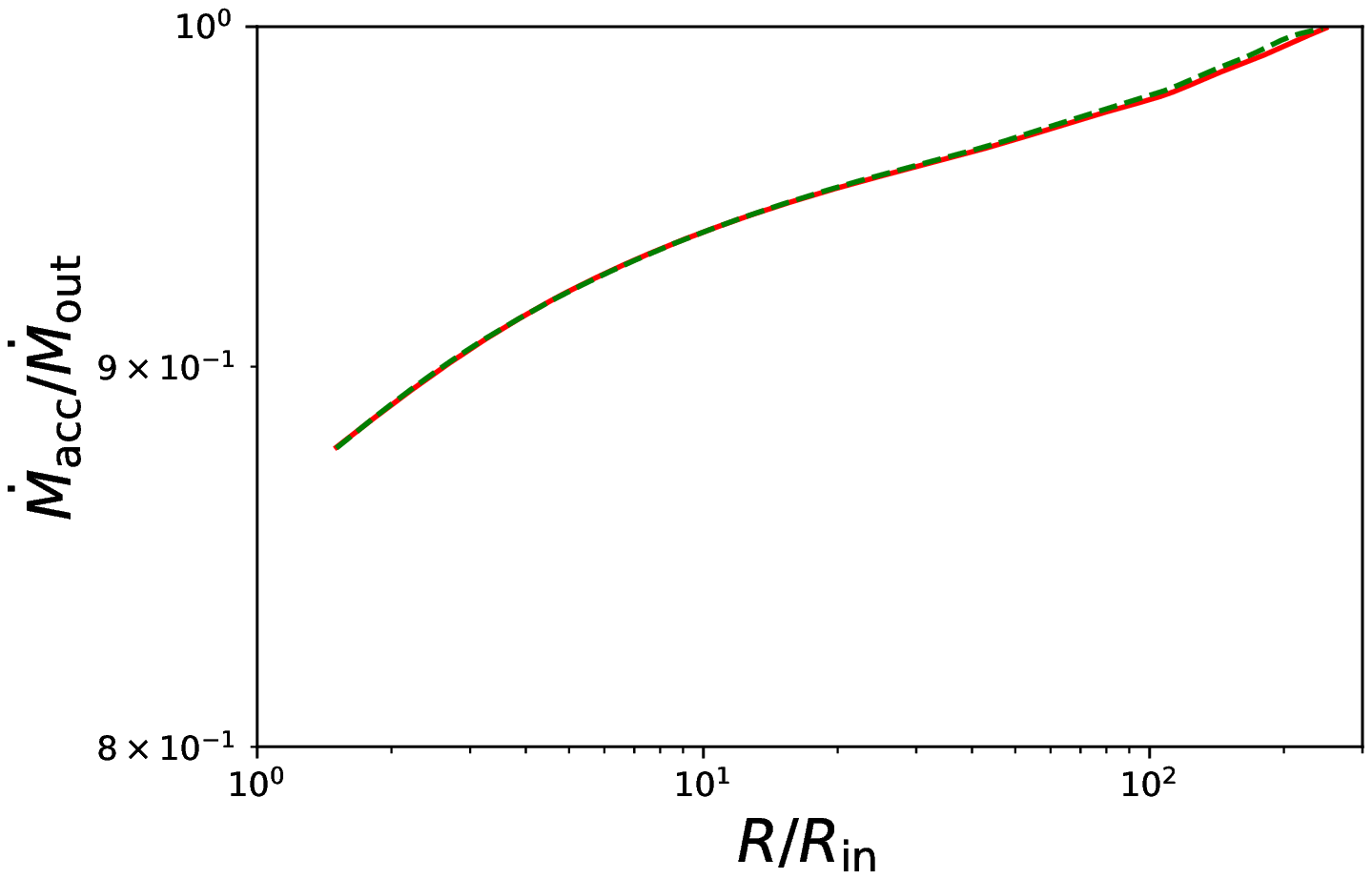}{0.5\textwidth}{}
}
    \caption{The same as Figure \ref{fig:vp1}, the red line shows the result for $\gamma=1.4$ while the green line is for $\gamma=5/3$. Left panel: solid line is labeled for poloidal velocity of the outflow $v_{\rm p}$, dotted line is for toroidal velocity $v_{\varphi}$. Right panel is the mass accretion rates in the disks.
    }
    \label{fig:vpvhi1_dmacc1}
\end{figure}

\section{Discussion}\label{sec:dissc}
As we know that the flux transport within a thin disk is very inefficient, a fast inward moving corona above a thin disk is expected to alleviate this dilemma \citep[][]{2012MNRAS.424.2097G,2009ApJ...701..885L,2021ApJ...909..158L}. The configurations of the large-scale poloidal magnetic field advected by a disk-corona system are {shown in Figure 1 of \cite{2021ApJ...909..158L}}, and we find that the field lines are significantly inclined to the surface of the corona, which implies that the gas in the corona will be easily driven into outflows by the field lines co-rotating with the gas in the disk-corona \citep[][]{1982MNRAS.199..883B}. The field configurations derived in our work are qualitatively consistent with those obtained in some previous numerical simulations \citep[][]{2018ApJ...857...34Z, 2020MNRAS.492.1855M}. Based on the configuration/strength of a large-scale magnetic field advected by a disk-corona system \citep[][]{2021ApJ...909..158L}, we calculate the dynamical properties of the outflows.

We find that the inclination angle of the field line with respect to the surface of the corona decreases with increasing radius. At the large radius, the effective potential barrier along the field line becomes shallow, which makes the slow sonic points shifting close to the surface of the corona. It leads to azimuthal field component being much stronger than the poloidal one, which may suffer non-axis-symmetric instabilities. The assumption of a potential field above the corona in this work is no longer valid, and the winds may probably be magnetically propelled circulation along the corona surface. A very similar case of the outflows magnetically driven from the outer region of the disk has been discussed in \citet{1994A&A...287...80C}. As we are more interested in the fast moving outflows from the inner region of the disk-corona, we will not be involved in such complexity of the winds driven from the outer region of the disk-corona.

 Due to a relatively larger mass loss rate in the outflow (i.e., a heavier outflow), it is therefore within a smaller radius the field line can enforce the gas in heavier outflow co-rotating with it, which leads to a smaller terminal velocity of the outflow. The toroidal field component is rapidly developed within a short distance from the field footpoint, as shown in the upper panel of Figure \ref{fig:2D3Dline}. Beyond the Alfv\'en surface the outflow is mainly driven by the magnetic gradient force due to the strong toroidal field. As mentioned above, all these two cases are calculated under the context of weak magnetic field, thus the field line can not enforce the gas in the outflow to co-rotate with the field line to a large radius where the toroidal component of the field dominates over the poloidal component. As a result, the position of the maximum toroidal velocity of the outflow along the field line is located slightly above the surface of the corona,  and it's value is close to local Keplerian velocity (see Figure \ref{fig:vphi}), which is near the Alfv\'en surface  (see the dotted lines in left panel of Figure \ref{fig:vpvphi_macc}). The properties of the magnetically driven outflows may also be affected by their radiation, which is beyond the scope of this work. Fortunately, the adiabatic index could describe its impact on the outflow dynamics to some extent. We re-calculate the whole problem with a different value of $\gamma=5/3$ for comparison, which are plotted
in Figures \ref{fig:vp1}-\ref{fig:vpvhi1_dmacc1}, as mentioned in Section \ref{sec:dynamics} we find that the results are not sensitive to the value of $\gamma$.

As we know that the dynamical structure of the disk, the large-scale field configuration of the disk, and the outflows are tightly coupled. To derive the dynamical structure of the disk, for instance, one needs to know the outflow properties first, while the calculations of the outflow solution require the specific field configuration/strength above the disk, which in turn decided by the whole structure of the disk itself \citep[see][for the details]{2019ApJ...872..149L}. In this work, the field advected by the corona is relative weak, which leads to outflows with low or moderate power. As we focus on the dynamical properties of the outflows, we neglect the impact of the outflows on the disk-corona structure, which is justified in part by the fact that only a small fraction of gas is driven into the outflows (see right panel of Figure \ref{fig:vpvphi_macc}).

The plasma $\beta$ parameter is important in deriving the properties of the outflow. In the field advection scenario of this work, the gas-to-magnetic ratio at the upper surface of the corona should be greater than unity, the field line is easily sheared into toroidal component in the outflow, and the resultant Alfv\'en surface is close to the surface of the corona. Thus, the outflows driven by the large-scale magnetic field advected by the corona are in general not very powerful (relative low mass loss rate and/or low terminal speed), which are consistent with the simple estimate in \citet{2018MNRAS.473.4268C}. \citet{2012ApJ...757...65S} also carried out time-dependent simulations on the outflows/jets launched magnetically from diffusive accretion disks, and they confirmed that the magnetocentrifugal acceleration is more efficient in a strong magnetic case (i.e., a lower plasma $\beta$), which may imply the outflow accelerated from a MAD (magnetically arrested disks) can reach a very high velocity with a low mass loss rate. Such  outflows may be able to account for the UFOs with mild relativistic velocities. In their work, the forces exerted on the outflow in the direction parallel and perpendicular to the field line are depicted, which clearly verify the relative importance of the forces in the acceleration process. However, we are not able to separate the forces (i.e., the gravity, pressure gradient, Lorentz, and centrifugal) exerted on the outflow in our calculations, as our calculations are carried out in the coordinates of the parallel and perpendicular to the field line, i.e., the force consists of the parallel and perpendicular components (see left panel of Figure \ref{fig:vpvphi_macc}), with which, in principle, the aforementioned forces can be derived, however, it would be very complicated. It would be worth trying to extract these forces information in our future work.

Although it is well believed that magnetic field must play an important role in the formation of the outflows/winds and the relativistic jets \citep[][]{1977MNRAS.179..433B,1982MNRAS.199..883B}, direct observation of the magnetic field is quite unlikely. Polarization observations of the emission from the relevant objects provide a powerful tool to measure the field configuration/strength, though it is usually somewhat model dependent. In this work, we have derived three-dimensional field configuration/strength (see upper panel of Figure \ref{fig:2D3Dline}), the density and velocity of the outflows, which can be used to calculated not only the emergent spectra of the disk-corona system but also the emergent spectrum of the outflows. Compared with the polarization observations, the derived results will surely help understand the physics of magnetic field origin in the disks and magnetically driven outflows. Recent high resolution polarization observations of M87 by the event horizon telescope (EHT) provided evidence of strong magnetic field near the black hole, i.e., the accretion flow around the black hole is likely to be in the MAD  state \citep[][]{2021ApJ...910L..13E}. Similar observations on the objects containing outflows magnetically driven from the disk-corona would be an ideal test of our model calculations, or the expansion of our calculations to the outflows driven from the MAD can be used to explain the observations of M87, which is beyond the scope of this work.  

In black hole accreting system, the hot corona is believed to be responsible for the non-thermal radiation in the X-ray bands \citep[][]{1993ApJ...413..507H}. Our calculations show that a fraction of the gas in the hot corona can be tapped into the outflows, which implies an outflowing corona located in the transition region of the corona to the outflow. Indeed, some early X-ray observations seem to require such outflowing coronas in the XRBs and AGNs \citep[e.g.,][]{1999ApJ...510L.123B,2014ApJ...783..106L}. Our model calculations provide a natural origin for such outflowing coronas. The further detailed calculations of the spectra of the disk-corona outflow system derived in this work would help understand the features of the X-ray observations in XRBs and AGNs, which will be reported in the future work.

%%%%%%%%%%%%%%%%%%%%%%%%%%%%%%%%%%%%%%%%%%%%%%%%
\acknowledgments
%%%%%%%%%%%%%%%%%%%%%%%%%%%%%%%%%%%%%%%%%%%%%%%%

We thank Xiaodong Duan and Weixiao Wang for the helpful discussions, and thank the referee for valuable comments/suggestions. This work is supported by the NSFC (11773050, 11833007, 12073023), the science
research grants from the China Manned Space Project with NO.
CMS-CSST-2021-A06, the CAS grant QYZDJ-SSWSYS023, and China Postdoctoral Science Foundation NO. 2021M702865.

%% For this sample we use BibTeX plus aasjournals.bst to generate the
%% the bibliography. The sample631.bib file was populated from ADS. To
%% get the citations to show in the compiled file do the following:
%%
%% pdflatex sample631.tex
%% bibtext sample631
%% pdflatex sample631.tex
%% pdflatex sample631.tex

\bibliography{ms}{}
\bibliographystyle{aasjournal}

%% This command is needed to show the entire author+affiliation list when
%% the collaboration and author truncation commands are used.  It has to
%% go at the end of the manuscript.
%\allauthors

%% Include this line if you are using the \added, \replaced, \deleted
%% commands to see a summary list of all changes at the end of the article.
%\listofchanges

\end{document}